\documentclass[11pt]{article}
\usepackage[english]{babel}
\usepackage{amssymb,amsthm}
\usepackage{amsmath}
\usepackage{float}
\usepackage[plainpages=false,hypertexnames=false,linktocpage=true]{hyperref}
\usepackage{tikz}
\usetikzlibrary{decorations.pathreplacing}
\usepackage{xcolor}
\colorlet{MyColorOne}{black!100}

\newcommand{\lightercolor}[3]{
    \colorlet{#3}{#1!#2!white}
}
\lightercolor{MyColorOne}{40}{MyColorOneLight}
\lightercolor{MyColorOne}{10}{MyColorOneLightest}
\lightercolor{MyColorOne}{5}{MyColorOneLightest2}
\usepackage[titletoc,toc,title]{appendix}
\usepackage{fullpage}
\newcommand{\Z}{\mathbb{Z}}
\newcommand{\ket}[1]{\mathopen| #1 \mathclose\rangle}
\newcommand{\vect}[1]{{\mathbf{#1}} }
\newcommand{\Vector}[1]{\boldsymbol{#1}}
\newcommand{\bd}{\begin{displaymath}}
\newcommand{\ed}{\end{displaymath}}
\newcommand{\be}{\begin{equation}}
\newcommand{\ee}{\end{equation}}
\newcommand{\bea}{\begin{eqnarray}}
\newcommand{\eea}{\end{eqnarray}}
\newcommand{\C}{\mathbb{C}}
\newcommand{\R}{\mathbb{R}}
\newcommand{\phic}{\theta_{\mbox{\tiny{crit}}}}
\newcommand{\pc}{p_{\mbox{\tiny{crit}}}}
\newcommand{\p}{\mathfrak{p}}
\newcommand{\e}{\epsilon}
\usepackage{wrapfig}
\usepackage{graphicx}
\usepackage[small,skip=0pt]{caption} 
\usepackage{subcaption}
\title{\bf Thermodynamics of Inozemtsev's Elliptic Spin Chain}
\date{}
\author{Rob Klabbers \\ 
\emph{II. Institut f\"{u}r Theoretische Physik, Universit\"{a}t Hamburg,}\\
\emph{Luruper Chaussee 149, 22761 Hamburg, Germany},\\[5mm] rob.klabbers@desy.de \\[5mm]
{\small This article is registered under preprint number: arXiv:1602.05133}}
\begin{document}
\begin{flushright}
    \footnotesize
    ZMP-HH/16-4
\end{flushright}
{\let\newpage\relax\maketitle} 
\thispagestyle{empty}


\begin{abstract}
We study the thermodynamic behaviour of Inozemtsev's long-range elliptic spin chain using the Bethe ansatz equations describing the spectrum of the model in the infinite-length limit. We classify all solutions of these equations in that limit and argue which of these solutions determine the spectrum in the thermodynamic limit. Interestingly, some of the solutions are not selfconjugate, which puts the model in sharp contrast to one of the model's limiting cases, the Heisenberg \textsc{xxx} spin chain. Invoking the string hypothesis we derive the thermodynamic Bethe ansatz equations (TBA-equations) from which we determine the Helmholtz free energy in thermodynamic equilibrium and derive the associated $Y$-system. We corroborate our results by comparing numerical solutions of the TBA-equations to a direct computation of the free energy for the finite-length hamiltonian. In addition we confirm numerically the interesting conjecture put forward by Finkel and Gonz{\'a}lez-L{\'o}pez that the original and supersymmetric versions of Inozemtsev's elliptic spin chain are equivalent in the thermodynamic limit. 
\end{abstract}
\newpage
\setcounter{tocdepth}{2}
\tableofcontents
\section{Introduction}
The Bethe ansatz has been one of the most powerful tools in the field of integrability in the past eighty years. Its origin dates back to Bethe's solution of the Heisenberg model for the ferromagnetic interaction of electrons from 1931 \cite{bethe}. Since then, analysis of numerous models other than spin chains benefited greatly from this ansatz, including the one-dimensional Bose gas \cite{LiebLin}, two-dimensional lattice models such as the six-vertex model \cite{PhysRev.162.162} and even $\mathcal{N}=4$ super Yang-Mills theory \cite{MinZar,SerbanStau}. Moreover, many extensions of Bethe ansatz have been found, including the thermodynamic Bethe ansatz \cite{Yang2,Takahashi01081971}, nested Bethe ansatz \cite{de1987finite} and asymptotic Bethe ansatz \cite{sutherland1971quantum,faddeev1979quantum,faddeev1996algebraic}. 
\\
\indent Heisenberg's spin-$1/2$ \textsc{xxx} spin chain is still ubiquitous in the research field centred around the Bethe ansatz. In an effort to generalize this spin chain, Inozemtsev proposed an elliptic spin chain characterized by the hamiltonian
\be
\label{ellH}
H = -\frac{J}{8} \sum_{\substack{j,k = 1 \\ j\neq k}}^L \wp_L(j-k)\left(\Vector{\sigma}_j\smash{\cdot}\Vector{\sigma}_k-1\right),
\ee
where $L$ is the number of sites of the spin chain, $J$ is the interaction parameter and $\wp_L$ is the Weierstra{\ss} elliptic function with periods $(L,i\pi/\kappa)$ (for $\kappa >0$) (see Appendix \ref{sec:elliptic}) and $\Vector{\sigma}$ is the usual vector of Pauli spin-$1/2$ operators \cite{Ino17}. Amazingly, this spin chain not only generalizes the Heisenberg \textsc{xxx} spin chain, which is recovered by taking $\kappa \rightarrow \infty$, but actually interpolates smoothly between the (nearest-neighbour) \textsc{xxx} spin chain and the long-range Haldane-Shastry spin chain, obtained in the limit $\kappa \rightarrow 0$. The Haldane-Shastry spin chain is solvable by exploiting its Yangian symmetry already present at finite length \cite{HSchainS,HSchainH}. Therefore, investigating Inozemtsev's elliptic spin chain may shed light on the relation between these two methods for finding exact solutions. In particular, the integrability of both the Heisenberg \textsc{xxx} spin chain and the Haldane-Shastry spin chain suggest that Inozemtsev's elliptic spin chain might also be integrable. Although a definite proof remains absent to date, research into this question has culminated in a proposed set of $L$ conserved quantities \cite{Ino32} and a description of eigenstates at finite and infinite $L$, which were found using an extended version of Bethe ansatz \cite{Ino21}. Another piece of evidence interestingly comes from the analysis of the level density of the spectrum of the spin chain, which agrees to great accuracy with some existing conjectures about chaotic versus integrable behaviour of quantum systems \cite{finkellopez1}. 
\\
\indent In fact, the spectrum of Inozemtsev's elliptic spin chain has been studied before. Dittrich and Inozemtsev probed the spectrum of Inozemtsev's infinite-length spin chain by classifying its two-particle bound states \cite{Dittrich}. Later, this spin chain was also used in a completely different context to calculate the first corrections to the dilatation operator in $\mathcal{N}=4$ super Yang-Mills theory. To this end asymptotic Bethe ansatz for Inozemtsev's spin chain was used to calculate corrections to the spectrum of the Heisenberg \textsc{xxx} spin chain as a truncated power series in $\kappa$ \cite{SerbanStau}, thereby providing some perturbative results on the spectrum of Inozemtsev's elliptic spin chain in the large volume limit. 
\\
\indent Finally, the related supersymmetric su$(1|1)$ version of Inozemtsev's elliptic spin chain was studied in \cite{finkellopez2} and shown to be integrable. It interpolates smoothly between the supersymmetrizations of the Heisenberg \textsc{xxx} spin chain (the \textsc{xx} spin chain at critical strength of the magnetic field) and of the Haldane-Shastry spin chain  \cite{sachdev2011quantum,haldane1994physics}. The thermodynamic limit of the su$(1|1)$ elliptic spin chain was studied and shown to correctly reproduce the behaviour of the aforementioned models in the appropriate limits. In addition, the Heisenberg \textsc{xxx} and Haldane-Shastry spin chain turn out to be equivalent to their supersymmetrizations in the thermodynamic limit and it has been hypothesized that this equivalence also carries over to the elliptic spin chain. 
\\[5mm]
In this work, we aim to gain additional information about the spectrum in the thermodynamic limit by invoking the string hypothesis \cite{bethe}, i.e. by assuming that the solutions of the Bethe ansatz equations in the infinite-length limit completely describe the thermodynamic behaviour of the model\footnote{In the paper \cite{Squeezed} a study of the thermodynamics of Inozemtsev's spin chain was announced, but to the author's best knowledge this study has never been published.}. After characterizing all the solutions, usually called \emph{strings}, we find integral equations describing the system in the thermodynamic limit. This method is quite standard for integrable models \cite{Takahashi01081971,Hubbard,arutyunov2009thermodynamic} and can be viewed as an extension of the method brought forward by Yang and Yang in \cite{Yang2}.\footnote{An application of the method by Yang and Yang to Inozemtsev's spin chain can be found in the author's \cite{MT}. In that unpublished work one can also find an account of part of the results discussed in the present work.} \\
\indent In Section $2$ we will recall the relevant models and point out some of their important properties. In Section $3$ we review the method of finding strings from a general perspective and apply it to the case of Inozemtsev's spin chain. In Section $4$ we give arguments why not all the found strings can be used to parametrize the spectrum and present a set of strings that should describe the thermodynamics of Inozemtsev's elliptic spin chain. In Section $5$ we apply the thermodynamic Bethe ansatz to these solutions to derive a set of integral equations that yield the free energy per site. In Section $6$ we compare a numerical solution of these equations to a direct computation of the free energy from the hamiltonian. Particular attention is paid to the relation to the Heisenberg \textsc{xxx} spin chain. We conclude in Section $7$ by summarizing our results. The appendices cover some basics on Weierstra{\ss} elliptic functions (Appendix \ref{sec:elliptic}), a thorough analysis of the important function $\phi$ (defined in equation \eqref{phi}) (Appendix \ref{sec:B}) and finally an analysis of convergence of solutions to the Bethe ansatz equations (Appendix \ref{sec:C}).
\section{Inozemtsev's elliptic spin chain}
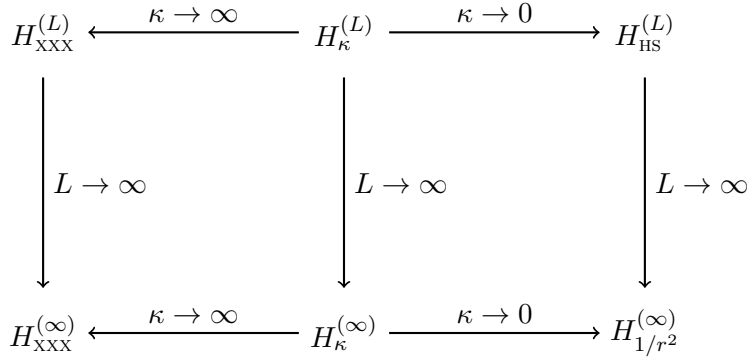
\begin{figure}[t!]
\centering
\begin{tikzpicture}
\def \a{0.6};
\node at (0,2) {$H_{\kappa}^{(L)}$};
\node at (0,-2) {$H_{\kappa}^{(\infty)}$};
\node at (4,2) {$H_{\textsc{hs}}^{(L)}$};
\node at (-4,2) {$H_{\textsc{xxx}}^{(L)}$};
\node at (-4,-2) {$H_{\textsc{xxx}}^{(\infty)}$};
\node at (4,-2) {$H_{1/r^2}^{(\infty)}$};
\foreach \x in {-1,1}
{
\draw[->,thick] (0-\a,2*\x)-- (-4+\a,2*\x);
\node[above] at (2,2*\x) {$\kappa \rightarrow 0$};
\draw[->,thick] (\a,2*\x)-- (4-\a,2*\x);
\node[above] at (-2,2*\x)  {$\kappa \rightarrow \infty$};
}
\foreach \x in {-1,...,1}
{
\draw[->,thick] (4*\x,2-\a)-- (4*\x,-2+\a);
\node[right] at (4*\x,0) {$L\rightarrow \infty$};
}
\end{tikzpicture}
\caption{A diagram showing the various limits to the hamiltonians of related models obtained from the hamiltonian of Inozemtsev's elliptic spin chain. In particular, the infinite-length Haldane-Shastry hamiltonian $H_{1/r^2}^{(\infty)}$ is of the form \eqref{genham} with potential $V_{1/r^2}^{(\infty)}(j) = 1/j^2$.}
\label{commut}
\end{figure}

\begin{figure}[t!]
\centering
\includegraphics[width=\textwidth]{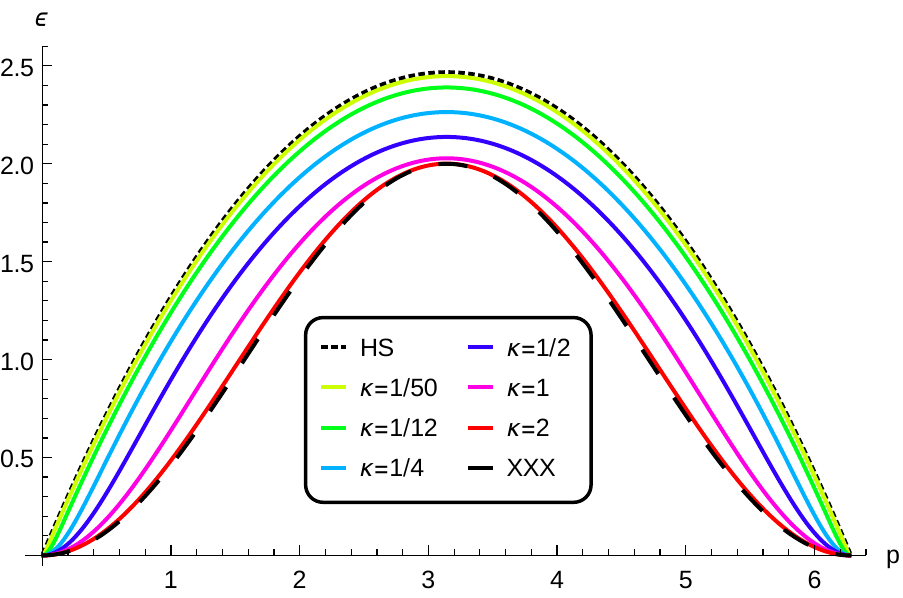}
\caption{(Colour online). The one-particle energies of the Haldane-Shastry spin chain $\e(p) = p(2\pi -p)/4$, Heisenberg \textsc{xxx} spin chain $\e(p) = 1-\cos(p)$ and Inozemtsev's spin chain see \eqref{1particle} for various $\kappa$ as a function of the quasi-momentum $p$.}
\label{energies}
\end{figure}

Inozemtsev's elliptic spin chain with spin $1/2$ as defined by the Hamiltonian in equation \eqref{ellH} has been studied extensively (see e.g. \cite{Ino17,Ino32,Ino_prf}). It is expected to be integrable, although this has not been completely proven since there is no proof that the found conserved quantities actually commute. There does exist a set of exact eigenfunctions in the form of a generalized Bethe ansatz and transcendental equations that determine the quasi-momenta. Various models can be reached starting from the elliptic spin chain by varying either the parameter $\kappa$ and/or the length $L$ of the chain. All of these spin-$1/2$ models are characterized by hamiltonians of the form
\be
\label{genham}
-\frac{J}{8} \sum_{\substack{j,k = 1 \\ j\neq k}}^{L} V(j-k)\left(\Vector{\sigma}_j\smash{\cdot}\Vector{\sigma}_k-1\right),
\ee 
where the potential $V$ can depend on the length $L$ of the chain, which is possibly infinite. In this work we will focus solely on the ferromagnetic case $J>0$. Following \cite{finkellopez2,Ino}, we accommodate these limits by redefining the hamiltonian \eqref{ellH} by rescaling and shifting the potential by site-independent factors: from now on we take Inozemtsev's elliptic spin chain to be defined by the hamiltonian 
\be
\label{inoresc}
H_{\kappa}^{(L)}=-\frac{J}{8} \sum_{\substack{j,k = 1 \\ j\neq k}}^{L} V_{\kappa}^{(L)}(j-k)\left(\Vector{\sigma}_j\smash{\cdot}\Vector{\sigma}_k-1\right),
\ee 
where 
\be
V_{\kappa}^{(L)}(j) = \frac{\sinh(\kappa)^2}{\kappa^2}\left(\wp_L(j) + \frac{2\kappa}{i \pi} \zeta_L\smash{\left( \frac{i \pi}{2\kappa}\right)} \right),
\ee
where $\zeta_L$ is the Weierstra{\ss} $\zeta$-function with quasi-periods $(L,i\pi/\kappa)$. If one sends $\kappa$ to infinity we reach the Heisenberg \textsc{xxx} spin chain (see \cite{Ino17} or Appendix A of \cite{finkellopez2}) with potential 
\be
\label{vxxx}
V_{\textsc{xxx}}^{(L)}(j) = \delta_{|j \mbox{ mod }L|,1}.
\ee
If one sends $\kappa$ to zero, one obtains the hamiltonian of the Haldane-Shastry (HS) spin chain with potential \cite{HSchainS,HSchainH}
\be
\label{HS}
V_{\textsc{hs}}^{(L)}(j) = \frac{\pi^2}{L^2 \sin^2  \frac{\pi j}{L}}.
\ee
On the other hand, if we keep $\kappa$ fixed and send $L\rightarrow \infty$ we reach Inozemtsev's infinite-length spin chain with potential
\be
\label{infIno}
V_{\kappa}^{(\infty)}(j) = \frac{\sinh^2\kappa}{\sinh^2 \kappa j}, 
\ee
which was treated extensively in \cite{Ino21}. All limits are summarized in Fig. \ref{commut}.
\\[5mm]
The exact solution of Inozemtsev's spin chain of infinite length is based on the su$(2)$-invariance of its local hamiltonians allowing for the $M$-particle ansatz
\begin{eqnarray}
\label{spinan}
\psi(n_1,\cdots,n_M)=
\prod_{1\leq \mu<\nu \leq M} \sinh^{-1}\kappa(n_{\mu}-n_{\nu}) \sum_{P\in S_M} (-1)^P\exp\left(\sum_{j=1}^M( ip_{Pj}-\kappa(M-1))n_j\right)\times \nonumber \\
 \sum_{\vect{m}\in W} c_{m_1\cdots m_M}(\vect{p})\exp\left( 2\kappa \sum_{j=1}^M m_{Pj}n_j \right), \nonumber
\end{eqnarray}
where $W$ denotes the set of all $\vect{m}\in \Z^M$ such that $0\leq m_i \leq M-1$ for all $1 \leq i\leq M$ and $S_M$ is the symmetric group of $M$ symbols. The coefficients $c_{m_1\cdots m_M}(\vect{p})$ can be found solving the set of equations 
\be
	\sum_{k \in \mathbb{Z}_{n_m,n_{m\smash{'}}}} c_{n_1,\cdots, n_m+k,\cdots,n_{m'}-k,\cdots,n_M} (\vect{p}) \left(n_m - n_{m'} + 2k + \frac{i}{2\kappa} (p_m -p_{m'}) \right) = 0,
\ee
where $\Z_{n,n'} = 
\{ k \in \mathbb{Z} \mid \max(-n,n'-M+1) \leq k \leq \min(M-1-n,n') \}$.
These eigenfunctions are closely related to the eigenfunctions of the continuous Calogero-Moser-Sutherland model with $1/\sinh^2$-interaction \cite{Ino21}. The associated eigenvalues are additive, the energy of an $M$-magnon state being given by 
\be
E_M(\vect{p}) = \sum_{i=1}^M \epsilon(p_i),
\ee
with (see also Fig. \ref{energies})
\bea
\label{1particle}
\epsilon(p)  &=& -\frac{J}{2}\sum_{\substack{n\in \Z \\ n\neq 0}} \frac{\sinh^2\kappa}{\sinh^2{\kappa n}}(\cos(p n ) - 1) \nonumber \\
&=&\frac{J\sinh^2\kappa}{2\kappa^2}\left(  -\frac{1}{2}\wp\smash{\left(\frac{ip}{2\kappa}\right)} +\frac{1}{2}\left( \frac{p}{\pi}\zeta\smash{\left(\frac{i\pi}{2\kappa}\right)}-\zeta\smash{\left(\frac{ip}{2\kappa}\right)}\right)^2 -\frac{2i\kappa}{\pi}\zeta\smash{\left(\frac{i\pi}{2\kappa}\right)}\right),
\eea
where the Weierstra{\ss} functions $\wp=\wp_1$ and $\zeta=\zeta_1$ are defined on the lattice with periods $(1,i \pi/\kappa)$.
Note that, unlike the finite-length case, solving the eigenvalue problem with this ansatz does not lead to any restrictions on the quasi-momenta and one needs to resort to other methods to find the spectrum of the model. A way to introduce Bethe equations is to follow the \emph{asymptotic Bethe ansatz} scheme (ABA), which can be summarized as imposing periodic boundary conditions on the asymptotic form of the eigenfunctions \eqref{spinan} \cite{Ino,Dittrich2}. This leads to Bethe equations (BE) (see \cite{Dittrich2})
\be
\label{BE}
e^{i p_j L} = \prod_{\substack{n=1 \\ n\neq j}}^M \frac{\phi(p_j) - \phi(p_n) +i}{\phi(p_j) - \phi(p_n) -i}, \qquad 1\leq j \leq M,
\ee
where $M$ denotes the total number of magnons and the meromorphic function $\phi$ is given by
\be
\label{phi}
\phi(p)=\frac{p}{2\pi i \kappa} \zeta\smash{\left( \frac{i\pi}{2\kappa}\right)} -\frac{1}{2i\kappa}\zeta\smash{\left( \frac{i p}{2\kappa}\right)}.
\ee
Solving these equations at $L\rightarrow \infty$ yields sets of quasi-momenta that are good candidates for parametrizing the spectrum of Inozemtsev's infinite-length spin chain, but one needs to to verify this by different means since usually the relation between quasi-momenta and physical states is not one-to-one. The solutions to \eqref{BE} might also be used to study the thermodynamic limit ($M,L\rightarrow\infty$ with $M/L$ fixed) of Inozemtsev's elliptic spin chain, since at very large $L$ the eigenfunctions of the elliptic spin chain can be approximated by those of the infinite spin chain, as was shown equivalently in the su$(1|1)$ case in \cite{finkellopez2}.\footnote{The hamiltonian of the supersymmetric models can be written as in \eqref{genham} but with $\Vector{\sigma}_j\smash{\cdot}\Vector{\sigma}_k$ replaced by the supersymmetric permutation operator $\mathcal{S}_{jk}$ acting as
\bd
\mathcal{S}_{jk} \ket{ s_1,\cdots,s_j,\cdots,s_k,\cdots,s_L} = (-1)^n \ket{ s_1,\cdots,s_k,\cdots,s_j,\cdots,s_L} 
\ed
with $n=s_j=s_k$ if $s_j=s_k$ and otherwise $n$ being the number of fermions on the sites $j+1,\cdots,k-1$.} \\
The system of equations \eqref{BE} is the usual form of BE, where for example the Bethe equations of the homogeneous Heisenberg \textsc{xxx} spin chain (BE$_{\textsc{xxx}}$) are of this form with the function $\phi$ replaced by 
\be
\phi_{\textsc{xxx}}(p) = \frac{1}{2}\cot\left(\frac{p}{2}\right),
\ee
although in that particular case $p$ can be replaced by $\phi_{\textsc{xxx}}$ in the BE$_{ \textsc{xxx} }$ altogether due to the form of $\phi_{\textsc{xxx}}$. Note that $\lim_{\kappa \rightarrow \infty} \phi = \phi_{\textsc{xxx}}$, implying that the BE$_{ \textsc{xxx} }$ can be found from equation \eqref{BE} by taking this limit. It is therefore natural to expect that the solutions to the BE \eqref{BE} with \eqref{phi} are closely related to the known results for the BE$_{ \textsc{xxx} }$. 
\section{Solving the Bethe equations asymptotically}
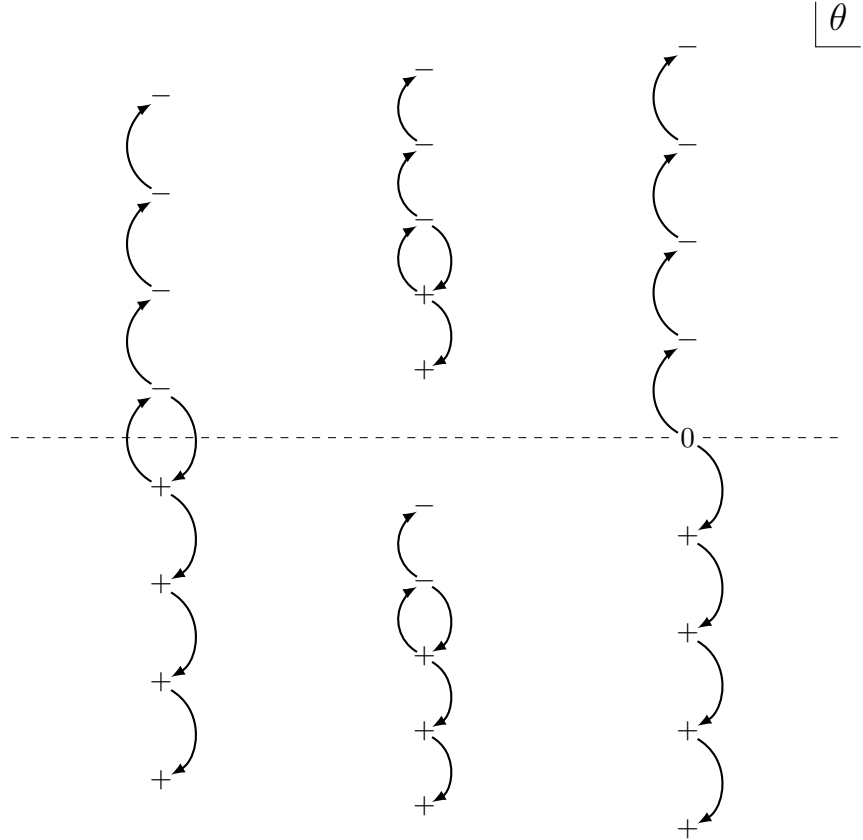
\begin{figure}[h!]
\begin{center}
\begin{tikzpicture}[>=latex,scale=1]
\node at (5.5,5.6) {\Large $\theta$};
\def \a{5.2};
\def \b{0.6};
\draw (\a,\a+\b) -- (\a,\a) -- (\a+\b,\a);
  \draw[ultra thin,dashed] (-5.5,0) -- (3.3,0);  	
  \draw[ultra thin,dashed] (3.7,0) -- (5.5,0);
\foreach \x in {1,...,2}
{
\node at ( 0, 1.9+ \x) {$-$};
\node at ( 0, -.1+ \x) {$+$};
\node at ( 0, -1.9- \x) {$+$};
\node at ( 0, .1- \x) {$-$};
\draw[->,thick](-0.1,0.95+\x) arc (240:120:0.5) ;
\draw[<-,thick](0.1, -0.05+\x) arc (300:420:0.5) ;
\draw[->,thick](-0.1,-0.85-\x) arc (240:120:0.5) ;
\draw[<-,thick](0.1, -1.85-\x) arc (300:420:0.5) ;
}
\node at (0,4.9) {$-$};
\node at (0,-4.9) {$+$};
\draw[->,thick](-0.1,3.95) arc (240:120:0.5) ;
\draw[<-,thick](0.1,-4.85) arc (300:420:0.5) ;

\node at (-3.5,0) {
\begin{tikzpicture}[>=latex,scale=1.3]
   	
\foreach \x in {1,...,4}
{
\node at ( 0, -4.5+ \x) {$+$};
\node at ( 0, -.5+ \x) {$-$};
}
\foreach \x in {1,...,4}
{
\draw[->,thick](-0.1,-1.45+\x) arc (240:120:0.5) ;
\draw[<-,thick](0.1, 0.55-\x) arc (300:420:0.5) ;
}
\end{tikzpicture}
};
\node at (3.5,0) {
\begin{tikzpicture}[>=latex,scale=1.3]

\foreach \x in {1,...,4}
{
\node at ( 0, \x) {$-$};
\node at ( 0, -\x) {$+$};
\draw[<-,thick](0.1, -4.95 +\x) arc (300:420:0.5) ;
\draw[->,thick](-0.1,-0.95+\x) arc (240:120:0.5) ;
}
\node at (0,0) {0};
\foreach \x in {1,...,4}
{
}
\end{tikzpicture}
};
\end{tikzpicture}
\end{center}
\caption{Sign configurations of minimal data in $\theta$-space. $\R\subset\C$ is indicated by the dashed line and the arrows indicate the structure that solves the BE: an arrow from sign $s_m$ to sign $s_n$ indicates that the BE with $j=n$ are satisfied because of the presence of $(\theta_m,s_m)$ on its right-hand side (so $(\theta_m,s_m)$ helps $(\theta_n,s_n)$). The left configuration is the standard string solution as the ones occurring for the \textsc{xxx} model. The middle configuration is a new feature of Inozemtsev's BE and consists of two connected components. For odd $M$, the allowed sets of minimal data look like the right configuration, with a real momentum in the middle, indicated by the $0$.}
\label{oplossingen}
\end{figure}

We are interested in solving the system of $M$ equations \eqref{BE} for an $M\in \mathbb{N}$ in the limit $L\rightarrow \infty$ for sets of noncoinciding\footnote{in accordance with the fact that the wavefunction parametrized by coinciding momenta vanishes.} complex momenta $\{p_j\}\in D$, with
\be
D= \{p\in \C| -\pi\leq\mbox{Re}(p) <\pi\}.
\ee
We can restrict $-\pi\leq$ Re$(p_j) <\pi$ for all $1\leq j \leq M$ due to the translation invariance of the spin chain. The total momentum and energy of these sets should be real, that is 
\bea
\sum_{j=1}^M p_j \in \mathbb{R},  \qquad \sum_{j=1}^M \epsilon(p_j) \in \mathbb{R}. 
\eea
The behaviour of the terms in \eqref{BE} in the limit $L\rightarrow \infty$ is quite simple: The left-hand side only depends on 
\be
s_j = \mbox{sign}(\mbox{Im}(p_j)),
\ee
with the sign function taking values in the set $\{+,0,-\}$, as follows: it
\begin{enumerate}
\item diverges if $s_j=-$,
\item converges to zero if $s_j=+$,
\item is of unit modulus if $s_j=0$.  
\end{enumerate}
On the other hand, we see that the right-hand side only depends on the images $\{\theta_j = \phi(p_j)\}$. This leads to the conclusion that to see whether a set of $\{p_j\}$ solves the BE, all we need to know is
\begin{enumerate}
\item the signs of the imaginary parts $\{s_j\}$ of the $\{p_j\}$,
\item the location of the images $\{\theta_j\}$.
\end{enumerate}
Depending on the exact form of $\phi$, this implies that different sets of momenta might correspond to a single set of \emph{minimal data} $\{(\theta_j,s_j)\}_{j\leq M}$ with a sign $s_j \in \{+,0,-\}$ to indicate the sign of the imaginary part of the associated momenta. To analyze the possible solutions as structured as possible, we will first characterize the allowed sets of  minimal data by the usual analysis for string solutions \cite{bethe,Hubbard}.
\\[5mm]
Consider a case in which $s_1 =+$. We see from the BE for $j=1$ that there must be an $n \leq M$ (and we can take $n=2$ without loss of generality) such that $\theta_1-\theta_2 \rightarrow -i$ as $L\rightarrow \infty$, such that also the right-hand side of the BE converges to zero. We will say colloquially that $(\theta_2,s_2)$ \emph{helps} $(\theta_1,s_1)$ to satisfy its BE. It means that in the limit the real parts of $\theta_1$ and $\theta_2$ coincide and that Im$(\theta_2) = $ Im$(\theta_1)+1$. There are three options for $s_2$. If $s_2=+$, the reasoning continues along the same line until we find an $s_j$ with either $s_j = -$ or $s_j=0$. We will not treat cases with $s_j=0$ here for simplicity, since they can be derived from our results without much work, but show a possible configuration on the right in Fig. \ref{oplossingen} nevertheless. If $s_2=-$, however, we see that the Bethe equation for $j=2$ is already satisfied due to the presence of $(\theta_1,s_1)$. Therefore, we do not need to add more tuples $(\theta_n,s_n)$ to the set to make it consistent (provided that the reality condition on momenta and energy are satisfied). By carrying out a similar reasoning for the case $s_1=-$, we see that the basic structure of a set of minimal data is a string of pluses and minuses as in Fig. \ref{oplossingen}.
From this analysis, we see that an allowed set of points $\{\theta_j\}_{j\leq M} $ should be a subset of
\be
\label{equidstrings}
\left\{\theta_R +\left(\theta_I-j+1\right)i \, | \, 1\leq j\leq m\right\}
\ee
for a certain $m\leq M$ and certain $\theta_{R,I} \in \mathbb{R}$. Note that we allow several $\theta$ to occupy the same point in $\theta$-space, as seems allowed by the above analysis: as long as there is another tuple $(\theta_n,s_n)$ that provides the correct limiting behaviour as $L\rightarrow \infty$, we can include any tuple we want. In particular, if $(\theta_2,s_2)$ is such that it helps $(\theta_1,s_1)$, it will also help any $(\theta_n,s_n)$ that satisfies $\theta_n=\theta_1$ and $s_n=s_1$. We will see in Section \ref{sec:treestructures} that as long as the basic structure is present, we can almost freely associate as many tuples as we want to a single point in $\theta$-space. These solutions cannot as easily be depicted as in Fig. \ref{oplossingen}. Of course, we are at the moment ignoring possible issues with convergence, which we will address in Section \ref{sec:convergence}. Also note that actual solutions to the BE should in the end have real momentum and energy. However, given a set of momentum associated to a set of minimal data $\{(\theta_j,s_j)\}_{j\leq M}$ as derived above, we can always add the complex conjugates of these momenta to the set to make sure that both total momentum and energy are real, as long as $\{\phi(p_j)\} \cap \{\overline{\phi(p_j)}\}$ is either empty or consists of one real element (see the middle and right configuration in Fig. \ref{oplossingen} respectively). This is possible due to the meromorphicity of $\phi$ and the one-particle energy $\epsilon$. 
\\[5mm]
The real question now is whether this very general analysis (and in fact more general than usually considered) is even necessary in the present case. Before going into details about this question, let us first make the connection with the known results for the \textsc{xxx} spin chain and see why we do not need this general approach in that case.
\subsection{Solutions for the \textsc{xxx} spin chain}
The Bethe equations for the Heisenberg \textsc{xxx} spin chain are 
\be
\label{XXXBE}
e^{ip_j L} = \prod_{\substack{n =1,\cdots,M \\  n\neq j}} \frac{\phi_{\textsc{xxx}}(p_j)-\phi_{\textsc{xxx}}(p_i)+i}{\phi_{\textsc{xxx}}(p_j)-\phi_{\textsc{xxx}}(p_i)-i}, \qquad 1 \leq j \leq M,
\ee
where $\phi_{\textsc{xxx}}(p) = \frac{1}{2}\cot\left( \frac{p}{2} \right)$, which is in this case usually called the rapidity function. The structure of the solutions to these equations is very simple. For each $M$, there exists a one-parameter family of string solutions of length $M$, which can be most conveniently parametrized in terms of the \emph{rapidities} $\lambda_j = 1/2\cot\left( p_j/2 \right)$ and is given by
\be
\label{XXXstrings}
\lambda_j = \lambda_R + 1/2(M+1-2j)i, \qquad \mbox{with} \, \lambda_R \in \mathbb{R}.
\ee
The reason for this simple structure is the bijectivity of $\phi_{\textsc{xxx}}$ as a function on D. Following the reasoning introduced in the previous section, we want to solve equations of the form $\phi_{\textsc{xxx}}(p_2) = \phi_{\textsc{xxx}}(p_1) \pm i$ (where $p_2$ is the unknown). By the bijectivity of $\phi_{\textsc{xxx}}$, these equations have \emph{unique} solutions, which leads to a unique set of momenta as soon as $p_1$ is fixed. Additionally, the sum of momenta must be real to ensure that the energy of the solution is real, which imposes that the rapidities have the prescribed imaginary parts given in \eqref{XXXstrings}. So due to the bijectivity of $\phi_{\textsc{xxx}}$, all the asymptotic solutions to the BE$_{\textsc{xxx}}$ are of the form as in \eqref{XXXstrings} and usually called \emph{string solutions}. This is no longer the case if the bijectivity of $\phi$ is lost, which turns out to be the case for Inozemtsev's spin chain. 
\subsection{Behaviour of $\phi$}
\label{sec:region}
\begin{wrapfigure}{r}{0.35\textwidth}
  \vspace{-20pt}
\begin{center}
\begin{tikzpicture}[>=latex,scale=0.7]
\def \a{0.1};
\path [fill=MyColorOneLightest] (-3,-2) rectangle (3,2);
\path [fill=MyColorOneLightest2] (-3,2) rectangle (3,6);
\path [fill=MyColorOneLightest2] (-3,-6) rectangle (3,-2);
\path [fill=MyColorOneLightest] (-3,6) rectangle (3,7);
\path [fill=MyColorOneLightest] (-3,-7) rectangle (3,-6);

\draw[ultra thick, MyColorOne] (-3,-2+\a) -- (-3,2-\a);
\draw[ultra thick, MyColorOne] (3,-2+\a) -- (3,2-\a);
\draw[ultra thick, MyColorOne] (-1,-2) -- (1,-2);
\draw[ultra thick, MyColorOne] (-3+\a,2) -- (-1,2);
\draw[ultra thick, MyColorOne] (3-\a,2) -- (1,2);

\draw[ultra thick, MyColorOneLight, dashed] (-3,2+\a) -- (-3,6-\a);
\draw[ultra thick, MyColorOneLight, dashed] (3,2+\a) -- (3,6-\a);
\draw[ultra thick, MyColorOneLight, dashed] (-1,2) -- (1,2);
\draw[ultra thick, MyColorOneLight, dashed] (-3+\a,6) -- (-1,6);
\draw[ultra thick, MyColorOneLight, dashed] (3-\a,6) -- (1,6);

\draw[ultra thick, MyColorOneLight, dashed] (-3,-2-\a) -- (-3,-6+\a);
\draw[ultra thick, MyColorOneLight, dashed] (3,-2-\a) -- (3,-6+\a);
\draw[ultra thick, MyColorOneLight, dashed] (-1,-6) -- (1,-6);
\draw[ultra thick, MyColorOneLight, dashed] (-3+\a,-2) -- (-1,-2);
\draw[ultra thick, MyColorOneLight, dashed] (3-\a,-2) -- (1,-2);

\draw[ultra thick, MyColorOne] (-1,6) -- (1,6);
\draw[ultra thick, MyColorOne] (-3+\a,-6) -- (-1,-6);
\draw[ultra thick, MyColorOne] (1,-6) -- (3-\a,-6);
\draw[ultra thick, MyColorOne] (-3,-6-\a) -- (-3,-7);
\draw[ultra thick, MyColorOne] (-3,6+\a) -- (-3,7);
\draw[ultra thick, MyColorOne] (3,-6-\a) -- (3,-7);
\draw[ultra thick, MyColorOne] (3,6+\a) -- (3,7);
\draw[thick,MyColorOne] (-3,2) circle [radius=\a];
\fill [white] (-3,2) circle (0.07);
\draw[thick,MyColorOne] (3,2) circle [radius=\a];
\fill [white] (3,2) circle (0.07);
\draw[thick,MyColorOne] (-3,-6) circle [radius=\a];
\fill [white] (-3,-6) circle (0.07);
\draw[thick,MyColorOne] (3,-6) circle [radius=\a];
\fill [white] (3,-6) circle (0.07);
\draw[thick,MyColorOneLight] (-3,6) circle [radius=\a];
\fill [white] (-3,6) circle (0.07);
\draw[thick,MyColorOneLight] (3,6) circle [radius=\a];
\fill [white] (3,6) circle (0.07);
\draw[thick,MyColorOneLight] (-3,-2) circle [radius=\a];
\fill [white] (-3,-2) circle (0.07);
\draw[thick,MyColorOneLight] (3,-2) circle [radius=\a];
\fill [white] (3,-2) circle (0.07);
\draw (0,-7) -- (0,7);
\draw (-4,0) -- (4,0);
\node[below] at (-0.2,0) {$0$};
\node[above] at (-3.5,0) {$-\pi$};
\node[above] at (3.3,0) {$\pi$};
\node[left] at (0,2.4) {$i\kappa$};
\node[left] at (0,-2.4) {$-i\kappa$};
\node[left] at (0,6.4) {$3i\kappa$};
\node[left] at (0,-6.4) {$-3i\kappa$};

\fill [MyColorOne] (1,2) circle (\a);
\fill [MyColorOne] (-1,2) circle (\a);
\fill [MyColorOne] (1,-6) circle (\a);
\fill [MyColorOne] (-1,-6) circle (\a);
\fill [MyColorOneLight] (1,6) circle (\a);
\fill [MyColorOneLight] (-1,6) circle (\a);
\fill [MyColorOneLight] (1,-2) circle (\a);
\fill [MyColorOneLight] (-1,-2) circle (\a);

\draw[thick] (-1,-0.1) -- (-1,0.1);
\draw[thick] (1,-0.1) -- (1,0.1);
\node[above] at (1,0) {$\pc$};
\node[above] at (-1,0) {$-\pc$};


\node at (1.5,1) {$D_f$};
\node[MyColorOneLight] at (1.5,4) {$D_{1}$};
\node[MyColorOneLight] at (1.5,-4) {$D_{-1}$};
\node at (1.5,6.5) {$D_{2}$};
\node at (1.5,-6.5) {$D_{-2}$};
\end{tikzpicture}
\end{center}
\caption{\small The complex strip $D$ and its partition into regions $D_i$.}
\label{region}
  \vspace{40pt}
\end{wrapfigure}
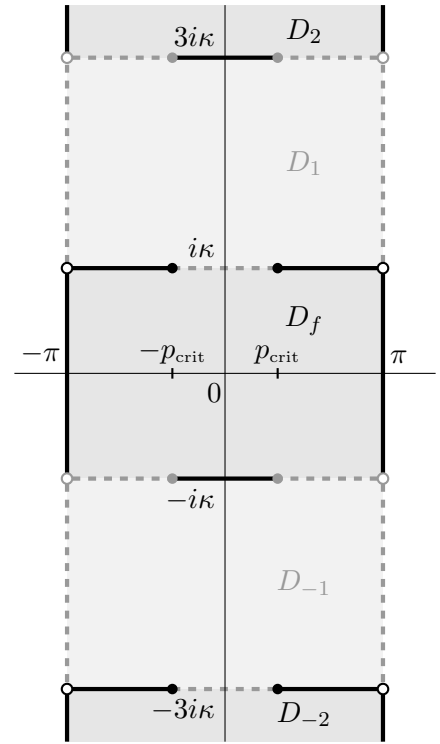
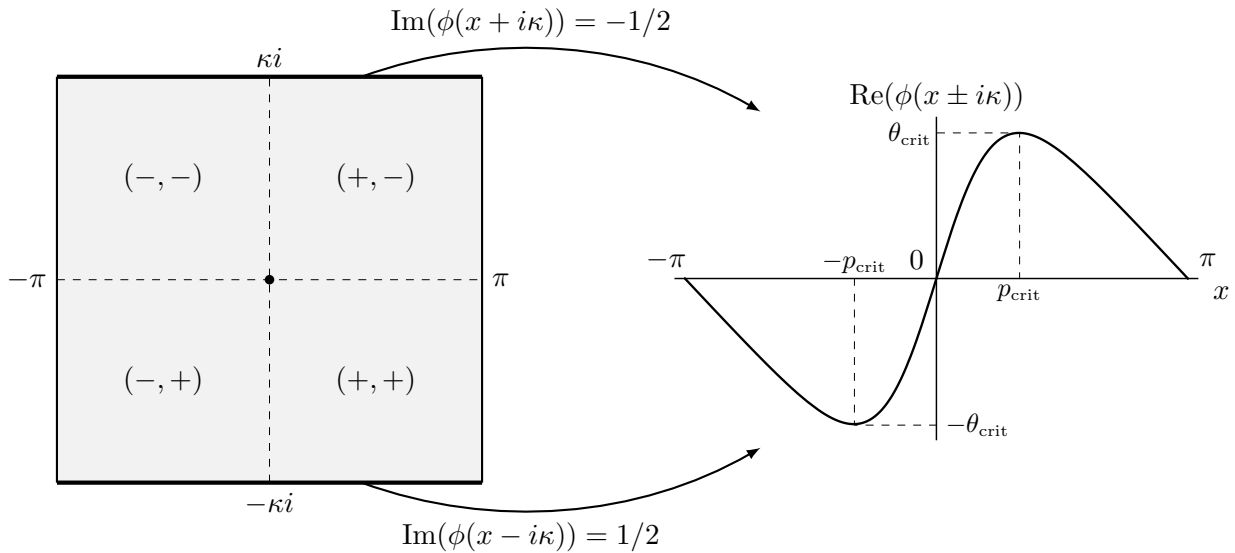
\begin{figure}[h!]
\begin{center}
\begin{tikzpicture}[>=latex,scale=0.9]
\path [fill=MyColorOneLightest2] (0,-3) rectangle (6.28,3);

\draw[ultra thick] (0,-3) -- (6.28,-3);
\draw[ultra thick] (0,3) -- (6.28,3);
\draw[thick] (0,-3) -- (0,3);
\draw[thick] (6.28,-3) -- (6.28,3);
\node[below] at (3.14,-3) {$-\kappa i$};
\node[above] at (3.14,3) {$\kappa i$};
\draw[ultra thin, dashed] (0,0) -- (6.28,0);
\draw[ultra thin, dashed] (3.14,-3) -- (3.14,3);
\node at (1.57,-1.5) {$(-,+)$};
\node at (3.14+1.57,-1.5) {$(+,+)$};
\node at (1.57,1.5) {$(-,-)$};
\node at (1.57+3.14,1.5) {$(+,-)$};
\fill [black] (3.14,0) circle (2pt);


\draw[->,thick](4.5, 3) arc (110:60:7);
\draw[->,thick](4.5, -3) arc (250:300:7);
\node at (7,3.8) {Im$(\phi(x+i\kappa)) = -1/2$};
\node at (7,-3.8) {Im$(\phi(x-i\kappa)) = 1/2$};

\node[left] at (0,0) {$-\pi$};
\node[right] at (6.28,0) {$\pi$};

\node at (13,0) {
\includegraphics[width=7cm]{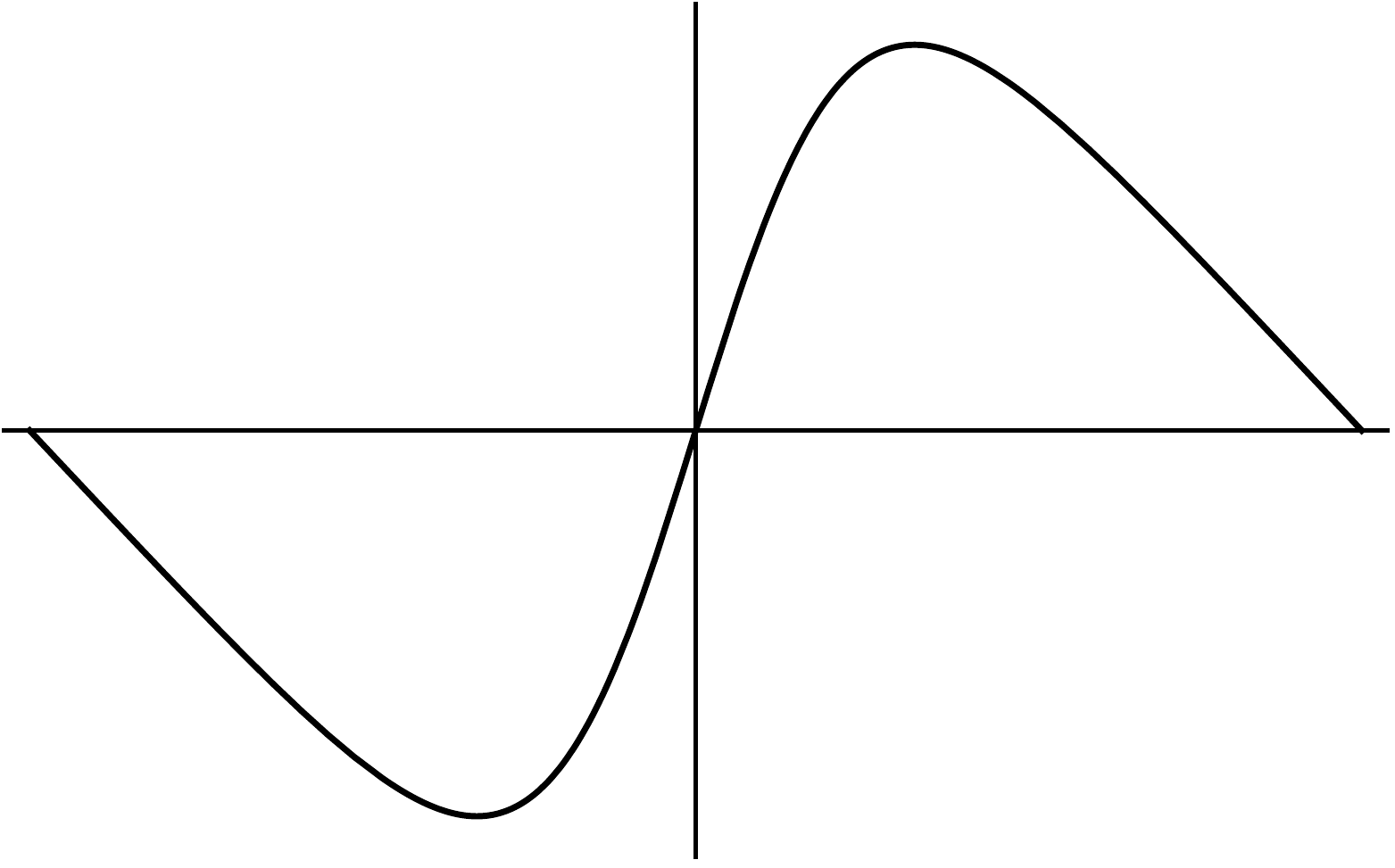}
};
\node at (9,0.3) {$-\pi$};
\node at (17,0.3) {$\pi$};
\node at (17.2,-0.2) {$x$};

\def \a{14.22};
\def \b{2.15};

\node at (12.6,\b) {\small $\phic$};
\node at (13.6,-\b) {\small $-\phic$};
\node at (13,2.7) {Re$(\phi(x\pm i\kappa))$};

\draw[dashed, ultra thin] (13,\b+0.02) -- (\a,\b+0.02) -- (\a,0.01);
\draw[dashed, ultra thin] (13,-\b) -- (26-\a,-\b) -- (26-\a,0.01);
\node at (12.7,0.3) {$0$};
\node at (\a,-0.2) {\small $\pc$};
\node at (26-\a,0.25) {\small $-\pc$};
\end{tikzpicture}
\end{center}
\caption{The range of $\phi$ on the domain $D_{{\scriptscriptstyle \leq} \kappa}$: On the left the signs in brackets indicate the sign of $(\mbox{Re}(\phi),\mbox{Im}(\phi))$ in that part of the domain. The black dot indicates the pole of $\phi$ at the origin. The behaviour of the real part of $\phi$ on the top and bottom domain boundary is explicitly shown in the plot on the right.}
\label{phibound}
\end{figure}

The function $\phi$ appearing in Inozemtsev's BE is odd and quasiperiodic, satisfying
\bd
\phi(p) = -\phi(-p), \quad \phi(p+2\pi) = \phi(p), \quad \phi(p+2i\kappa) = \phi(p) - i,
\ed
which means that its behaviour on the region 
\be
\label{Dsmallkappa}
D_{\scalebox{0.5}{$\leq$} \kappa} = \{ p\in D | \mbox{Im}(p) \leq \kappa \}
\ee
completely determines its behaviour on $D$. One can prove using the argument principle that $\phi: D_{\scalebox{0.5}{$\leq$} \kappa}\rightarrow \C$ is almost bijective.\footnote{With \emph{almost bijective} we mean that there exists a restriction of $\phi$ to a domain differing from $D_{\scalebox{0.5}{$\leq$} \kappa}$ by a set of measure zero that is bijective.}\footnote{This is shown in Appendix \ref{sec:B}.} $\phi$ is certainly surjective, but it attains twice those $\theta \in \C$ for which Im$(\theta) = \pm 1/2$ and $-\phic < |$Re$(\theta)|< \phic$, where $\phic>0$ depends on the parameter $\kappa$ and is defined by 
\be
\phic = \mbox{Re}\left(\phi(\pc +i\kappa)\right),
\ee
where $\pc$ is the unique solution on $[0,\pi]$ to the equation
\bea
\frac{d}{dp} \phi(p+i \kappa) = 0.
\eea
The preimages of these $\theta$'s lie on the top and bottom boundary of $D_{\scalebox{0.5}{$\leq$} \kappa}$, i.e. where Im$(p) = \pm \kappa$. This behaviour is illustrated in Figure \ref{phibound}. 
\\[5mm]
The quasi-periodicity and almost bijectivity of $\phi$ when restricted to $D_{\scalebox{0.5}{$\leq$} \kappa}$ inspires to introduce a partition of $D$ into regions such that $\phi$ is bijective when restricted to such a region: the \emph{fundamental region} $D_f$ is defined as
\be
D_f = \{ p \in D \,  |  \,0 \leq | \mbox{Im}(p)| <\kappa \} \cup \{q + \kappa i \in D\,  |\,\pi>| q| \geq \pc \} \cup \{q - \kappa i \in D \, | \, | q| <\pc \}
\ee
and the region $D_n$ is defined as the region obtained by shifting $D_f$ by $2\kappa i n$, that is $D_n = D_f + 2\kappa i n$, as can be seen in Fig. \ref{region}. The partition $\{D_n\}_{n\in \Z}$ of $D$ is such that the restrictions $\phi\big|_n$ to $D_n$ are bijective functions onto $\C$. This will make it easier to categorize the momentum sets that belong to a certain set of minimal data $\{(\theta_j, s_j)\}$. Finally, note that there exists exactly one other partition consisting of connected sets that differs from ours, which can be created by mirroring this partition in the real line. 
\subsection{Solutions for Inozemtsev's spin chain}
The fact that $\phi$ is so far from being injective has great consequences for the solutions of the BE of Inozemtsev's spin chain. Equations of the form 
\be
\label{eqn1}
\phi(p) = \theta
\ee
for a given $\theta \in \C$, have an countably infinite set of solutions, parametrized by the region index in which each of the solutions lies. In particular, the equation \eqref{eqn1} has solutions for $p$ with positive and with negative imaginary parts. This makes it possible for a string solution to consist of two parts, as in the middle of Figure \ref{oplossingen}: each part in itself forms a consistent solution to the BE, but only the sum of the parts has real energy and momentum. These new solutions also have more degrees of freedom than the usual string solutions: whereas the usual string solutions (such as the left configuration in Fig. \ref{oplossingen}) have no freedom in choosing the imaginary parts of the $\theta$'s, the new solutions can be shifted in the imaginary direction as long as the two parts remain complex conjugate and distinct. More precisely, for $m$ distinct $\theta_j$ we can choose $\theta_I$ parametrizing the imaginary part of the $\theta_j$ as in  \eqref{equidstrings} to be anything from the set
\be
\mathbb{R}_m:=\mathbb{R}\setminus\left\{0,\frac{1}{2},1,\frac{3}{2},\cdots, m+1\right\}.
\ee
As an example, consider the solution consisting of the four momenta 
\bd
\{p_1=0.108+4.62 i,\, p_2=0.280-0.659i,\, p_3 = \overline{p_1} = 0.108-4.62 i,\,p_4=\overline{p_2} =  0.280+0.659i\},
\ed
for the case where $\kappa =1.26$. It consists of two connected components and has $m=M$ (i.e. non-coinciding $\theta_j$), $\theta_I = 1.8$ and $\theta_R = 0.6$. This is just one of the countably infinite number of solutions specified by these $\theta_R,\theta_I$: there are infinitely many $D_n$ from which $p_1$ can be chosen and the same is true for $p_2$. We see that solutions with $m=M$ are not a one-parameter family (as was the case for the \textsc{xxx} spin chain); there are $2$ continuous parameters and $M/2$ discrete ones needed to specify an $M$-momenta solution of this type. 
\subsubsection{Solutions with coinciding $\theta_j$}
\label{sec:treestructures}
\begin{figure}[h!]
\begin{center}
\begin{tikzpicture}[>=latex,scale=1.2]
\node at (1.5,2) {$-$};
\node at (2.5,2) {$-$};
\node at (3.5,2) {$\cdots$};
\node at (4.5,2) {$-$};

\node at (-1,1) {$+$};
\node at (0,1) {$+$};
\node at (1.,1) {$\cdots$};
\node at (2,1) {$+$};
\node at (3,1) {$-$};
\node at (4,1) {$-$};
\node at (5,1) {$\cdots$};
\node at (6,1) {$-$};
\node at (2.5,-0.3) {$\vdots$};
\node at (2.5,0.3) {$\vdots$};
\node at (-1,-1) {$+$};
\node at (0,-1) {$+$};
\node at (1,-1) {$\cdots$};
\node at (2,-1) {$+$};
\node at (3,-1) {$-$};
\node at (4,-1) {$-$};
\node at (5,-1) {$\cdots$};
\node at (6,-1) {$-$};

\node at (1.5,-2) {$+$};
\node at (2.5,-2) {$+$};
\node at (3.5,-2) {$\cdots$};
\node at (4.5,-2) {$+$};

\node at (-3,2) {$M_1$};
\node at (-3,1) {$P_{2},M_{2}$};
\node at (-3,-0.3) {$\vdots$};
\node at (-3,0.3) {$\vdots$};
\node at (-3,-1) {$P_{m-1},M_{m-1}$};
\node at (-3,-2) {$P_m$};
\end{tikzpicture}
\end{center}
\caption{\small The sign configuration of a set of coincident minimal data. The $P_j,M_j$ indicate the number of pluses and minuses at each level.}
\label{tree1}
\end{figure}
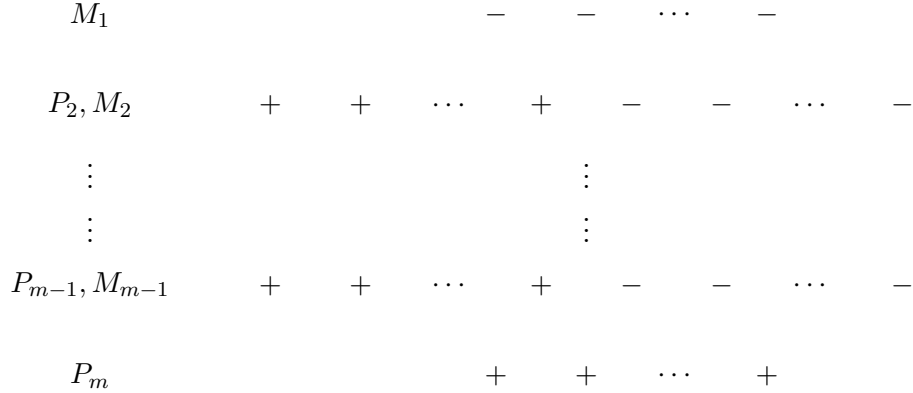

\begin{figure}[h]
\begin{center}
\begin{subfigure}{6cm}
\centering
\begin{tikzpicture}[>=latex,scale=1]
\node at (4,0) {$-$};
\node at (4,-1) {$+$};
\node at (3,-1) {$+$};
\node at (5,-1) {$+$};

\draw[->,thick](3.0, -0.8) arc (160:120:1.41);
\draw[->,thick](5, -0.8) arc (20:60:1.41);

\draw[<-,thick](4.9, -0.9) arc (230:210:3);
\draw[<-,thick](3.1, -0.9) arc (310:330:3);

\draw[<-,thick](4.05, -0.8) arc (330:390:0.6);
\draw[->,thick](3.95,-0.8) arc (210:150:0.6);

\end{tikzpicture}
\caption{}
\end{subfigure}
\begin{subfigure}{6cm}
\centering
\begin{tikzpicture}
\node at (7,0) {$+$};
\node at (9,0) {$-$};
\node at (8,-1) {$-$};
\node at (9,-2) {$+$};
\node at (7,-2) {$+$};
\node at (8,1) {$-$};

\draw[->,thick](7, 0.2) arc (160:120:1.41);
\draw[->,thick](8.9, 0.15) arc (20:60:1.41);
\draw[<-,thick](7.1, 0.1) arc (310:330:3);
\draw[->,thick](8.14,-0.85) arc (290:345:1.15);

\draw[->,thick](7.0, -1.8) arc (160:120:1.41);
\draw[->,thick](9, -1.8) arc (20:60:1.41);
\draw[<-,thick](8.9, -1.9) arc (230:210:3);
\draw[<-,thick](7.1, -1.9) arc (310:330:3);
\end{tikzpicture}
\caption{}
\end{subfigure}
\end{center}
\caption{\small Two examples of sign configurations parametrizing a coinciding solution. As in Fig. \ref{tree1} the arrows indicate which signs are helped by which others.}
\label{tree01}
\end{figure}
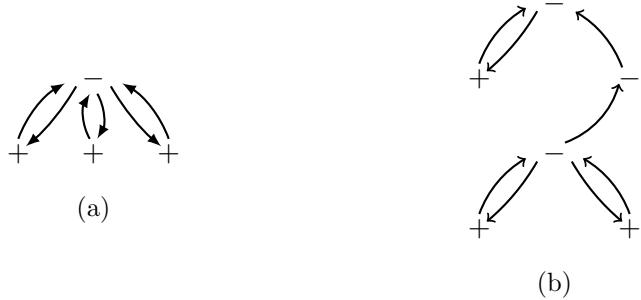

Although the sets of minimal data considered in the previous section are already an extension to the usual string analysis, Inozemtsev's BE allow even more general sets. The fact that we are only interested sets of non-coinciding momenta does not mean that also the set of $\theta_j$ should be non-coinciding. The non-injectivity of $\phi$ precisely allows us to associate any number of momenta to any particular value $\theta$. Moreover, in many cases we can associate momenta with both positive and negative imaginary part to each value. In order to be able to characterize these sets of minimal data more easily, we will no longer allow the $\theta_j$ to be coinciding, but instead associate multiple signs $s_{j,{i_j}}$ to a single $\theta_j$, that is we rewrite minimal data 
\be
\{(\theta_j,s_j)\}_{j\leq M} \rightarrow \{ (\theta_j,(s_{j,{1}},\cdots,s_{j,{l_j}}))\}_{j\leq M}, 
\ee
where now $\theta_j=\theta_n$ implies $j=n$. We can depict these sets of minimal data, which we will call \emph{coincident minimal data}, by placing the $s_{j,{i_j}}$ belonging to the same $\theta_j$ on a horizontal line (the \emph{level}). In this way, we can depict a set of coincident minimal data as has been done in Figure \ref{tree1}, where level $j$ contains $P_j$ pluses and $M_j$ minuses and the total number of levels is $m\leq M$. These numbers satisfy 
\be
\sum_{j=1}^m l_j =\sum_{j=1}^m \left(P_j +M_j\right) = M.
\ee
Thus in Figure \ref{tree1}, there are $M_1$ momenta with negative imaginary part associated to the image point $\theta_R +\theta_I i$ with biggest imaginary part, $M_2$ momenta with negative imaginary part to the image point $\theta_R +(\theta_I-1) i$ and $P_2$ momenta with positive imaginary part, etc. 
In this configuration, a sign $s_{j,{i_j}} = +$ on level $j$ receives help from all the $s_{j+1,{i_{j+1}}}$, whereas a sign $s_{j,{i_j}} = -$ receives help from all the $s_{j-1,{i_{j-1}}}$. 
Sets of coincident minimal data can be parametrized as follows: we fix an integer $M$ and an $m\leq M$ and choose an allowed sign configuration conform Figure \ref{tree1}.\footnote{Allowed sign configurations must have an arrow pointing towards each of its signs, indicating that each BE associated to a particular sign indeed has a term on its right-hand side such that its limiting behaviour as $L\rightarrow \infty$ is correct. Loosely speaking, we can enumerate the options by choosing a partition of $M$ using $2M$ non-negative integers, but this slightly overcounts the number of options.} Then we choose $\theta_R\in \mathbb{R}$ and $\theta_I \in \mathbb{R}_M$. An actual solution to the BE, a \emph{coinciding solution}, also requires the choice of a region $D_n$ for each of the $M$ signs in our configuration. Generically, there is an infinite number of allowed regions. It is clear that these solutions enjoy even more freedom than the ones considered in the previous sections. 
\\[5mm]
However, one might argue that for coinciding solutions we can no longer follow the naive construction of string solutions, since in this case taking the limit $L\rightarrow \infty$ becomes more problematic; indeed this is the case, because many different terms might converge or diverge on the right-hand side of a given BE for one of the momenta of a coinciding solution. In Appendix \ref{sec:C} we address this question more carefully, but we will see in the rest of our analysis that the question whether or not one can still take the limit is irrelevant.
\\[5mm]
{\bf Example.} Let us give some examples of possible coincident solutions of this type. Two examples of configurations are depicted in Figure \ref{tree01}. To find momentum sets corresponding to these configurations, we set $\kappa=1.26$ and $\theta_R=1.4$ arbitrarily. 
\\For example (a), we have $M=4$ and $m=2$. We can choose $\theta_I\in \mathbb{R}_4$, so let us pick $\theta_I=1.89$ arbitrarily. We choose regions $D_1,D_2,D_3$ for the plus signs and region $D_{0}$ for the minus sign. We must add the complex conjugates to let the solution have real momentum and energy and we end up with
\bd
\left\{ 0.244 + 2.175i, 0.132 +4.761i, 0.080 + 7.3334i, 0.244 - 0.345 i \right\} + \mbox{complex conjugates},
\ed
with energy $E_8 =-1.57234$ (again $J=1$).
\\[5mm]
For example (b), we have $M=6$ and $m=4$. We can also choose $\theta_I\in \mathbb{R}_M$ and we pick $\phi_I = 2.6$. For the lower two plus signs we use regions $D_f$ and $D_1$ and for the one on level $2$ we choose region $D_1$ as well. We use region $D_f$ for the momenta of all the minus signs. We again have to add complex conjugates to end up with a solution with real total momentum and energy. The solution is
\bea
\{0.687+0.213 i, 0.618 + 2.232i, 0.156 + 2.222i, 0.618-0.288 i, 0.300 - 0.361i, 0.156-0.298i\} \nonumber \\ +\mbox{complex conjugates}, \nonumber
\eea
with energy $E_8 = 0.211$. As shown in Appendix \ref{sec:C}, this solution does have a defect: there is no consistent way to consider the limit $L\rightarrow \infty$ for this solution, implying it is not a good candidate to parametrize the spectrum. 
\section{Pruning the solution set}
\label{sec:pruning}
The solutions presented in the previous sections obey the rules that are usually obeyed by string solutions of Bethe equations, such as the ones for the \textsc{xxx} model. Some of the features of the new solutions do raise questions: the solutions have too many degrees of freedom to execute the usual string hypothesis program, i.e. assume that the string solutions accurately describe the thermodynamic behaviour of the model and derive thermodynamic Bethe ansatz equations. In particular, excluding momenta in favour of $\theta$'s would mean having to introduce an uncountably infinite number of types of particles. We might expect, however, that these issues are due to the fact that only a subset of our set of solutions contains information about the spectrum of Inozemtsev's spin chain and most of the solutions to the BE are actually non-physical: they have some sort of defect that forces us to discard them as physical solutions. 
\\[5mm]
This is indeed the case, there are four main types of defects to be found in our set of solutions:
\begin{enumerate}
\item the associated wavefunction does not vanish at infinity, that is the momenta do not parametrize a bound state. 
\item they do not correspond to a string solution to the Heisenberg \textsc{xxx} spin chain in the limit $\kappa\rightarrow \infty$.
\item there is no consistent way to consider the limit $L\rightarrow \infty$. 
\item the associated wavefunction is identically zero.
\end{enumerate}
In the following sections we will consider these defects and discard the solutions that suffer from these defects. 
\subsection{Non-vanishing wavefunctions}
We consider the case of a two-particle solution to the BE, which induces a wavefunction parametrized by $p_1,p_2 \in D_{\pm i}$, where without loss of generality we can assume that Im$(p_1-p_2)<0$. A simple argument shows that this wavefunction does not vanish at infinity for $i\geq2$: the amplitude of the wavefunction is given by \cite{Ino_prf}
\be
|\psi(n_1,n_2)|^2 = 4 |\sinh^{-2} \kappa(n_1-n_2)| \left( e^{2\kappa(n_1-n_2)} +e^{-2\kappa(n_1-n_2)} - e^{i(n_1-n_2)(p_1-p_2)}- e^{-i(n_1-n_2)(p_1-p_2)} \right)
\ee
and, since Im$(p_1-p_2)<0$, we see that this only tends to zero in the limit $|n_1-n_2| \rightarrow \infty$ if $|$Im$(p_1)|\leq\kappa$, i.e. if $p_{1,2} \,\smash{\in} \, D_{{\scriptscriptstyle \leq} \kappa}$. Thus two-particle bound states must have all their momenta in $D_{\scalebox{0.5}{$\leq$} \kappa}$, which contains $D_f$ and part of the boundaries of $D_{\pm 1}$. 
\\[5mm]
Unfortunately, the complicated form of the wavefunctions for $M>2$ makes it difficult to prove a similar statement for bound states consisting of more than $2$ particles, although numerical analysis of the wavefunctions shows that the statement seems to be true at least up to $M=6$.
\subsection{Relation to the Heisenberg \textsc{xxx} spin chain}
To get more evidence for the fact that the solutions with momenta outside $D_{\scalebox{0.5}{$\leq$} \kappa}$ are non-physical, we take a closer look at the relationship between Inozemtsev's infinite spin chain and the \textsc{xxx} spin chain. Let us therefore first consider a general solution to Inozemtsev's BE: it consists of a set of $\theta_j$, an assignment of momenta to each of the $\theta_j$ and the regions where each of these momenta can be found. If we take the limit towards the \textsc{xxx} spin chain ($\kappa \rightarrow \infty$), all the momenta that do not lie in $D_{\scalebox{0.5}{$\leq$}\kappa}$ acquire infinite imaginary part, since they lie on the outside of $D_{\scalebox{0.5}{$\leq$} \kappa}$ which fills up $D$ entirely in this limit. Comparing this to the allowed string solutions for the Heisenberg \textsc{xxx} chain \eqref{XXXstrings} and stipulating that all solutions should flow to a \textsc{xxx} string in the limit $\kappa \rightarrow \infty$ also suggests we should abandon solutions that have momenta outside of $D_{\scalebox{0.5}{$\leq$} \kappa}$. 
\subsection{Convergence issues}
\label{sec:convergence}
The arguments above potentially reduce the solution set enormously, but their origin lies in applying restrictions that are not related to the solving of the BE itself. Interestingly, there is a subtle issue arising because some solutions have coinciding $\theta_j$, which forces us to look more carefully at the procedure of taking the limit $L\rightarrow \infty$ when we are looking at the BE. For a standard string solution, there are usually only two terms on the right-hand side of the BE of $p_j$ that do not have a finite limiting value, but converge to zero or diverge. One of the two terms is there to make sure that the equation is satisfied in the $L\rightarrow \infty$ limit, but the other one actually counteracts this. By associating to each momentum a speed with which its limiting value is reached, one can take the limit in a consistent way. For solutions with coinciding $\theta_j$ however, the case is more complicated, because more terms influence the limit. In Appendix \ref{sec:C}, we discuss this matter in detail and show that there is indeed a consistent way to consider the limit for standard string solutions and also give an example of a coinciding solution for which there is not. This excludes some of the coinciding solutions from being in the spectrum, but most of them remain to be candidates. In particular, we cannot exclude more solutions than can already be be excluded using the previous two arguments. 
\subsection{Vanishing wavefunctions}
\label{sec:exceptions}
If we restrict the domain of our momenta to $D_{\scalebox{0.5}{$\leq$} \kappa}$, we have almost completely excluded the possibility of coinciding $\theta_j$. In fact, the only remaining solutions of this kind must be built up from momenta living on the boundary of $D_{\scalebox{0.5}{$\leq$} \kappa}$, since only on the boundary is $\phi$ still non-injective (see Fig. \ref{phibound}). So for $|\theta_R| <\phic$ there are $2$ solutions of the equation
\be
\phi(p+i\kappa) = \theta_R-i/2,
\ee
which we name $p_1$ and $p_2$. In this way we can build four two-particle bound states, by pairing the momenta as follows:
\bea
\label{2part}
\{p_1 +i\kappa, p_1-i\kappa\}, \qquad \{p_2 +i\kappa, p_2-i\kappa\}, \nonumber \\
\{p_2 +i\kappa, p_1-i\kappa\}, \qquad \{p_1 +i\kappa, p_2-i\kappa\}.
\eea
The bound states on the first line are of the form $p_1=p-i\kappa$, $p_2=\overline{p_1}$. A simple computation shows that the wavefunctions of these bound states vanish identically. Numerical analysis up to $M=6$ suggests that this holds in general for wavefunctions parametrized by a set of momenta for which $p_n = p_m +2\kappa i$ for some $n,m$. If we assume this is indeed true, we can no longer built coincident solutions. However, we can still build two types of particles out of momenta in the region $D_{\scalebox{0.5}{$\leq$} \kappa}$ for $|\phi_R|<\phic$ and even $M$: there are two two-particle bound states (see the lower line of equation \eqref{2part}) that one can use as a basis for building a solution, after which there is no choice for the remaining particles, since they are fixed by the requirement of real energy. However, one can check that these two types of bound states have exactly the same energy and total momentum and in fact parametrize the exact same wavefunction; they parametrize the same bound state. This situation is very similar to the one encountered in \cite{arutyunov2007string} and our solution is the same: we only keep one of the two. Although the choice is arbitrary, we can choose by restricting the allowed momenta domain even further, from $D_{\scalebox{0.5}{$\leq$} \kappa}$ to $D_f$. Choosing the other bound state amounts to partitioning the domain $D$ in the alternative way as described in Section \ref{sec:region}. 
\\[5mm]
Note that the two-particle bound states we have found here are actually very peculiar: they are not self-conjugate when viewed in momentum space, which is a novel feature of Inozemtsev's spin chain (see also \cite{Dittrich}), but the total momentum and energy of this bound state are real. In fact, this is even more interesting when one realizes that the self-conjugacy of the string solutions 
\begin{wrapfigure}{r}{0.4\textwidth}
\vspace{-10pt}
\begin{tikzpicture}
\node at (0,0) {
\includegraphics[width=6cm]{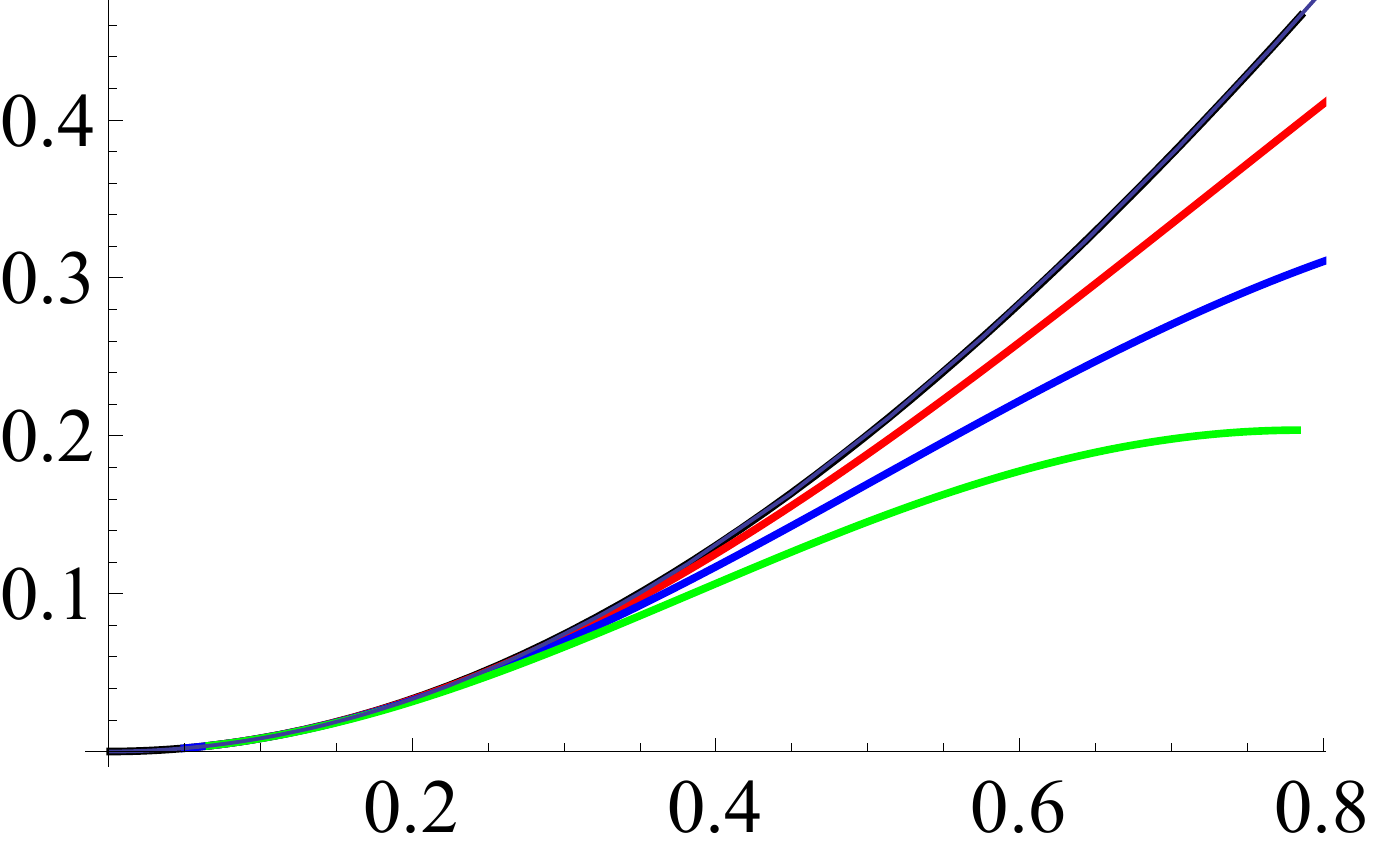}
};
\node at (2.6,-2.3) {$p\rightarrow$ };
\node at (-2.7,2.3) {$\tilde{E_i}(p)$};
\node at (3.2,1.6) {$\tilde{E_2}$};
\node at (3.2,0) {$\tilde{E_4}$};
\node at (3.2,2.1) {$\tilde{E_1}$};
\node at (3.2,0.8) {$\tilde{E_3}$};
\end{tikzpicture}
\vspace{-15pt}
\caption{(Colour online). The rescaled energies $\tilde{E_i}$ of bound states consisting of up to $4$ particles for $\kappa=1.23$ and $J=1$. }
\vspace{-20pt}
\label{bound-state energy}
\end{wrapfigure}
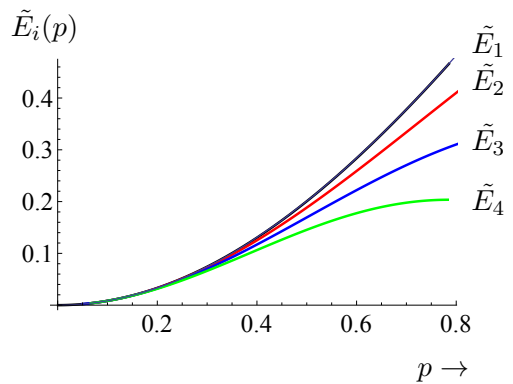
in the spectrum of the Heisenberg \textsc{xxx} model can be traced back to the underlying algebraic structure \cite{vladimirov1986proof}. Despite this difference, we will see in the next section that the equations describing the thermodynamic behaviour of Inozemtsev's spin chain bear a striking resemblance to those for the \textsc{xxx} model. 

\subsection{Remaining solutions}
All of the arguments presented above indicate that we should only consider solutions built up out of momenta from the fundamental region $D_f$. As another check to see that we are on the right track we have plotted in Fig. \ref{bound-state energy} for $J=1$ the rescaled energies 
\be
\tilde{E_i}(p) = \frac{E_M(Mp)}{M},
\ee
where $E_M$ is the energy of an $M$-string lying in the fundamental region as a function of total momentum. We see clearly that the inequality
\be
ME_1(p) \geq E_M(Mp)
\ee
is satisfied for all plotted $M$ and numerical analysis shows this is true at least up to $M=40$. This implies that the solutions are indeed bound states for positive $J$, because the energies of these states are smaller than the sum of one-particle energies. The next step now is to invoke the string hypothesis and assume that the remaining solutions accurately describe the thermodynamic behaviour of Inozemtsev's spin chains. That will allow us to perform the thermodynamic Bethe ansatz program to derive equations that describe the free energy of Inozemtsev's spin chains, which we will do in the next section. 
\section{Thermodynamic Bethe ansatz}
Since we are interested in the thermodynamic regime of Inozemtsev's elliptic spin chain, we want to send the number of quasi-particles $M$ and the length of the chain $L$ to infinity, while keeping the ratio $M/L$ fixed. The well-known method we will deploy here to take this limit for the BE \eqref{BE} is called thermodynamic Bethe ansatz (TBA) and the resulting set of equations are usually called TBA-equations.
\\[5mm]
Let us first summarize which string solutions we consider to be relevant in the thermodynamic limit: since we now have a one-to-one relation between $p$'s and $\theta$'s, we can no longer create solutions with coinciding $\theta_j$, but also lost our freedom to choose $\theta_I$. Our restricted $\phi$ is meromorphic and bijective, implying we can no longer make strings like the one portrayed in the middle of Fig. \ref{oplossingen}, because for all $p \in D_f$ we have that sign$($Im$(p))= -$sign$($Im$(\phi(p))$. Therefore, the remaining strings are of the form
\be
\left\{\theta + \left(j-\frac{M+1}{2}\right) i \, | \, 1\leq j \leq M, \theta \in \R \right\},
\ee
which we will call \emph{$Q$-strings}. Their total momentum and energy are given by 
\bea
\p_{Q}(\theta) &=& \sum_{j=1}^Q \phi^{-1}\left( \theta + \left(j-\frac{Q+1}{2}\right) i \right) \nonumber \\
E_Q(\theta) &=& \sum_{j=1}^Q \e\left(\phi^{-1}\left( \theta + \left(j-\frac{Q+1}{2}\right) i \right)\right),
\eea
where $\e$ is the one-particle energy given in \eqref{1particle} and unfortunately no explicit formula for $\phi^{-1}$ is known. As any momentum set $\{p_j\}$ must be built up from bound states, we can fuse the BE for the composite particles parametrized by our $Q$-strings:
\be
\label{fused}
e^{i \p_{P}\left(\theta_{P,l}\right) L } = \prod_{Q=1}^{\infty} \prod_{r=1}^{N_Q}S_{PQ}\left(\theta_{P,l},\theta_{Q,r}\right),
\ee
where the $N_Q\geq 0$ denotes the number of $Q$-strings and where
\bea
S_{PQ}(\theta,\theta')&=& \prod_{j=1}^P \prod_{k=1}^Q S\left(\theta+\left(j-\frac{P+1}{2}\right) i ,\theta'+\left(k-\frac{Q+1}{2}\right) i \right) \nonumber \\
&=& \prod_{j=1}^P \prod_{k=1}^Q \frac{\theta-\theta'-(\frac{P+Q}{2} -1)i+\left(j-k\right) i}{\theta-\theta'-(\frac{P+Q}{2} +1)i+\left(j-k\right) i},
\eea
with
\be
S(\theta,\theta') = \frac{\theta - \theta' +i}{\theta - \theta' -i}.
\ee
Taking logarithms in \eqref{fused} we get
\be
c_P(\theta_{P,l}) L = I_{P,l},
\ee
where the $I_{P,l}$ are integer quantum numbers and 
\be
\label{counting}
c_{P}(\theta) L = \frac{\p_{P}\left(\theta\right)}{2\pi} L -\frac{1}{2\pi i}\sum_{Q} \sum_{r=1}^{N_Q}\log S_{PQ}\left(\theta,\theta_{Q,r}\right)
\ee
are the counting functions. We can check numerically that these functions are monotonically increasing. Now we can introduce particle and hole densities $\rho_Q,\bar{\rho}_Q$ that should satisfy 
\be
\label{holedens}
\rho_Q(\theta) + \bar{\rho}_Q(\theta) = \frac{d c_Q}{d\theta}(\theta).
\ee
In the limit $L\rightarrow \infty$ the counting functions get transformed as the summations become integrals:
\bea
\frac{1}{2\pi i}\frac{1}{L}\sum_{Q} \sum_{r=1}^{N_Q}\log S_{PQ}\left(\theta,\theta_{Q,r}\right) \rightarrow \frac{1}{2\pi i}\sum_{Q} \int_{-\pi}^{\pi} d\theta' \log S_{PQ}\left(\theta,\theta'\right)\rho_Q(\theta'), 
\eea
We define the convolution
\be
f\star g(\theta) = \int_{\R} d\theta' f(\theta-\theta')g(\theta').
\ee
Taking the derivative explicitly, we see that \eqref{holedens} becomes
\be
\label{rep}
\rho_P(\theta) + \bar{\rho}_P(\theta) = \frac{1}{2\pi}\frac{d\p_{P}\left(\theta\right)}{d\theta}  -\sum_{Q} K_{PQ} \star \rho_Q (\theta),
\ee
where we have used the kernels
\bea
K_P(\theta) =  \frac{1}{\pi } \frac{P}{P^2+\theta^2}  \mbox{ for } P\geq 1 \mbox{ and } K_0(\theta) = \delta(\theta) \nonumber \\
K_{PQ}(\theta) = K_{|P-Q|} + K_{P+Q} +2\sum_{j=1}^{\min(P,Q)-1} K_{|P-Q|+2j}.
\eea
Note that these kernels are exactly the same as those appearing in the derivation of the TBA-equations for the \textsc{xxx} spin chain. In fact, our entire derivation differs from that one only because our formulae for $\p_Q$ and $E_Q$ cannot be written in terms of elementary functions. By varying \eqref{rep}, we get
\be
\label{variation}
\delta\rho_P + \delta\bar{\rho}_P = -K_{PQ} \star \delta\rho_Q,
\ee
where we sum over the repeated indices. Now we can introduce a free energy density and find the point of thermodynamic equilibrium:
\be
\label{fe1}
f=e-Ts,
\ee
where
\be
e = \sum_{Q}  \int_{\R} d\theta E_Q(\theta) \rho_Q(\theta).
\ee
The entropy density is defined as
\be
s=\sum_{Q} \int_{\R} d\theta\left( (\rho_Q + \bar{\rho}_Q)\log(\rho_Q+\bar{\rho}_Q) - \rho_Q\log \rho_Q - \bar{\rho}_Q \log \bar{\rho}_Q \right).
\ee
Varying, substituting equation \eqref{variation} and changing integration variables in the kernel term we end up with
\be
0=\delta f = \sum_{Q} \int_{\R} d\theta \left(E_Q(\theta) -T\left(\log \frac{\bar{\rho}_Q}{\rho_Q}(\theta) - \sum_{P} K_{QP} \star \log \left(1+ \frac{\rho_P}{\bar{\rho}_P}(\theta)\right) \right) \right)  \delta \rho_Q(\theta).
\ee
This directly leads to the TBA-equations of Inozemtsev's spin chain, which when we introduce the $Y$-functions  $Y_Q = \frac{\bar{\rho}_Q}{\rho_Q}$ read
\be
\label{TBA}
\log Y_Q = \frac{E_Q}{T} + \sum_{P=1}^{\infty} K_{QP}\star\log \left(1+ 1/Y_P\right),
\ee
which are of exactly the same form as those for the Heisenberg \textsc{xxx} model; the only difference sits in the definition of the energies $E_Q$. Moreover, after sending $\kappa\rightarrow \infty$ we recover the TBA-equations for the \textsc{xxx} spin chain. 
\subsection{Free energy}
Much of the information about the system in thermal equilibrium can be extracted from the density of the Helmholtz free energy $f$. We can express the free energy solely in terms of $Y$-functions as follows: plugging in our definition of $Y$'s into our definition of $f$ \eqref{fe1} leads to
\bea
f&=& \sum_{Q=1}^{\infty} \int_{\R} d\theta \left(E_Q\rho_Q -T\left( (\rho_Q\log(1+Y_Q) + \bar{\rho}_Q\log(1+(Y_Q)^{-1})\right)\right).
\eea
Now, using equation \eqref{variation} we can replace $\bar{\rho_Q}$ by $\rho_Q$:
\bea
f&=&T \sum_{Q=1}^{\infty} \int_{\R} d\theta \left(\rho_Q(\theta)\left(E_Q/T -\log Y_Q\right) -\frac{1}{2\pi}\frac{d\p_{Q}}{d\theta}\left(\theta\right)\log(1+(Y_Q)^{-1}) \right) \nonumber \\
&+& \sum_{Q=1}^{\infty} \int_{\R} d\theta \sum_{Q=1}^{\infty}  \left(K_{QP} \star\log\left(1+(Y_P)^{-1}\right)\right)(\theta) \rho_Q(\theta).
\eea
Using the fact the the $Y$-functions should obey the TBA-equations \eqref{TBA} we are left with the expression
\be
f=-\frac{T}{2\pi} \sum_{Q=1}^{\infty} \int_{\R} d\theta \frac{d\p_{Q}}{d\theta}\left(\theta\right)\log(1+1/Y_Q(\theta)).
\ee
In the next section, we will use this expression for the free energy to compare our equations describing the thermodynamics of the model with a more straightforward method starting from the hamiltonian. However, let us for completeness first mention how one could simplify the TBA-equations further. 
\subsection{$Y$-system}
One can find a simpler-looking set of equations using the function 
\be
s(\theta) = \frac{1}{4\cosh\left( \frac{\pi \theta}{2}\right)}
\ee
together with its pseudoinverse $s^{-1}$ defined by
\be 
s^{-1} \star f (x) = \lim_{\delta\rightarrow 0} \left( f(x+i-i\delta) + f(x-i +i \delta)\right)
\ee
and the property that 
\be
s\star \left(K_{P-1}+K_{P+1}\right) =  K_P  \mbox{ for } P\geq 1. 
\ee
These equations, the \emph{$Y$-system for Inozemtsev's spin chain}, take the following form:
\bea
\label{Ysystem}
Y_1^+ Y_1^- &=& \exp\left({1/T \left(\left(\hat{K_1}^{-1} + K_1\right)\star E_1  - E_2\right)}\right)(1+Y_2)\nonumber \\
Y_M^+ Y_M^- &=& \exp\left({1/T \left(\left(\hat{K_1}^{-1} + K_1\right)\star E_M - E_{M-1} - E_{M+1}\right)}\right) (1+Y_{M-1})(1+Y_{M+1}),
\eea
with $M>1$ in the second line and where the superscripts $\pm$ indicate shifts of $\pm i$ and the inverse $\hat{K_1}^{-1}$ means that 
\be
\hat{K_1}^{-1}\star K_1 (x) = \delta (x).
\ee
The $Y$-system of Inozemtsev's spin chain reduces to the $Y$-system for the \textsc{xxx} spin chain in the by now well known limit $\kappa \rightarrow \infty$ given by
\bea
\label{YsystemXXX}
Y_1^+ Y_1^- &=& (1+Y_2)\nonumber \\
Y_M^+ Y_M^- &=& (1+Y_{M-1})(1+Y_{M+1}), \quad M\geq 2
\eea
but for finite $\kappa$ the exponential prefactors do not seem to simplify. 
\\[5mm]
The $Y$-system looks simpler than the TBA-equations, but also has a downside: it admits more solutions than the ones we are interested in alone. One has to complement the $Y$-system with a set of asymptotics and possibly other analytic data to find the solution describing the thermodynamics of Inozemtsev's spin chain. Therefore we will later use the TBA- equations \eqref{TBA} to find numerical results. One important piece of data one can extract from the $Y$-system is the value of the $Y$-functions as $T\rightarrow \infty$: in this limit the exponential prefactors become unity, leaving us with the $Y$-system \eqref{YsystemXXX}. From \eqref{TBA} we see that in this limit the $Y$-functions should become constant functions and plugging this information into the $Y$-system gives the asymptotic result
\be
Y_M^{T\rightarrow \infty} = M(M+2),
\ee
which one can use to solve the TBA-equations numerically. 
\section{Solving the TBA-equations numerically}
Obtaining numerical results\footnote{All the numerical analysis was done in Mathematica 10.0.} for Inozemtsev's spin chains using the TBA-equations \eqref{TBA} is interesting for several reasons: firstly, we can use numerical results to check our equations, thereby implicitly corroborating the usage of the string hypothesis, used to the TBA-equations, as well as checking our treatment of the solutions of Inozemtsev's BE in Section \ref{sec:pruning}. Secondly, with numerical results we can check the hypothesis put forward in \cite{finkellopez2} that the normal and supersymmetric versions of Inozemtsev's spin chains are equivalent in the thermodynamic limit. \\
\indent To perform these checks we have gathered three types of numerical data: (1) the free energy for the \textsc{xxx} and Inozemtsev's spin chains from TBA-equations\footnote{The Heisenberg \textsc{xxx} TBA-equations and their derivation can be found e.g. in \cite{takahashi2005thermodynamics}.}, (2) the free energy for the \textsc{xxx}, Haldane-Shastry and Inozemtsev's spin chains computed from the finite-length hamiltonians and (3) the free energy for the supersymmetric versions of these three types of spin chains. 
\subsection{Methods}
(1) Solving the TBA-equations can be done quite fast using Fast Fourier Transform to compute the convolutions in a Picard iteration scheme. To be able to this we have cut the number of $Y$-functions to not more than $35$ and treated the real line as a grid of (typically) $2^8,2^9$ points. In our particular case the tricky part is computing the energies $E_Q$ efficiently, because their definitions contain the inverse of $\phi$, for which no explicit formula exists. We have written a program that is capable of finding inverses numerically, but finding inverses is still a time-consuming task compared to performing the iterations. \\
\indent The iteration scheme takes a set of $Y$-functions and iterates using the TBA-equations until stability is reached, that is until the biggest pointwise difference between ingoing and outgoing $Y$-functions is smaller than $10^{-10}$. Then the number of $Y$-functions is increased and a stable solution of this set of $Y$-functions is found. For both sets of $Y$-functions the free energy is calculated and if the relative difference between the found free energies is smaller than $10^{-5}$ J we declare the solution stable and otherwise keep increasing the number of $Y$-functions. This approach is very similar to the one used in e.g. \cite{okiji1984thermodynamic,usuki1990thermodynamic} and following these authors we believe that our results should be accurate at least up to a few percent. In particular, we have also explicitly computed the free energy from the TBA-equations for the Heisenberg \textsc{xxx} spin chain. 
\\[5mm]
(2) We have also compared the results from the TBA-equations with another calculation of the free energy per site: given a spin chain hamiltonian $H$ for finite $L$, one straightforwardly derives that 
\be
\label{matrixformula}
f_{\text{Tr}} = -\frac{T}{L} \log \mbox{Tr} \exp( -H/T),
\ee
where $L$ is the length of the spin chain and the subscript reminds us how we calculated this free energy. As long as $L$ is not too large (typically $L\leq 15$) one can perform these matrix operations explicitly reasonably fast. We have done this for the \textsc{xxx}, HS and Inozemtsev's spin chains, using the finite-length hamiltonians with potentials \eqref{vxxx},\eqref{HS} and \eqref{inoresc} respectively for increasing $L$ until the results stabilized (in this case that means that consecutive terms differ by less than $1-2\%$). Extrapolating the relative difference suggests that the results are also accurate up to a 5 percent. 
\\[5mm]
(3) Finally, we have reproduced the results in \cite{finkellopez2}, giving the free energy per site of the supersymmetric versions of the \textsc{xxx}, HS and Inozemtsev's spin chains. Computing the relevant integrals numerically was done with a high degree of precision (up to $50$ digits). 
\subsection{Results}
\begin{figure}[t!]
\centering
\includegraphics[width=\textwidth]{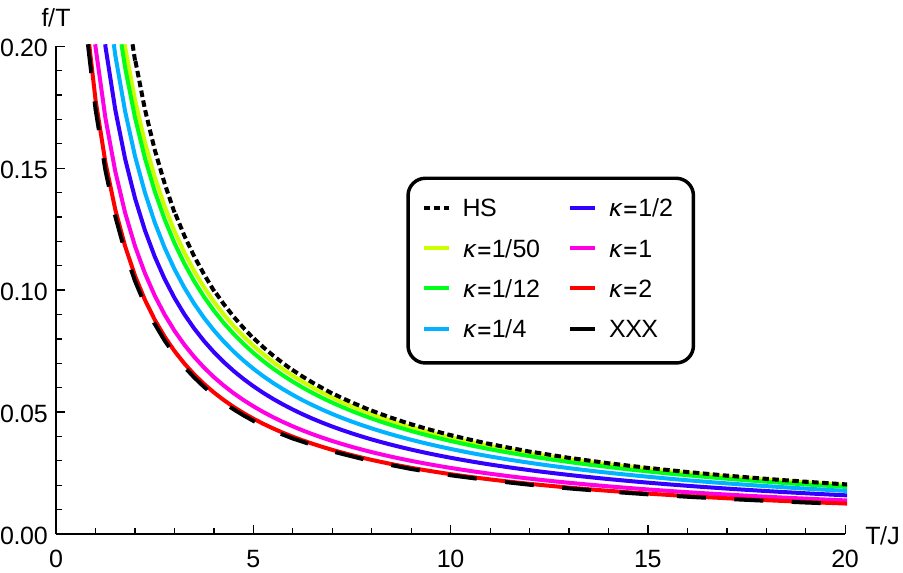}
\caption{(Colour online). The free energies of Inozemtsev's spin chain for different $\kappa$ as calculated from the TBA-equations \eqref{TBA}, along with the free energy for the \textsc{xxx} spin chain as calculated from TBA and the free energy of the HS model as calculated using \eqref{matrixformula}.}
\label{InotoXXX}
\end{figure}

\begin{figure}
\centering
\begin{subfigure}{8cm}
\includegraphics[width=8cm]{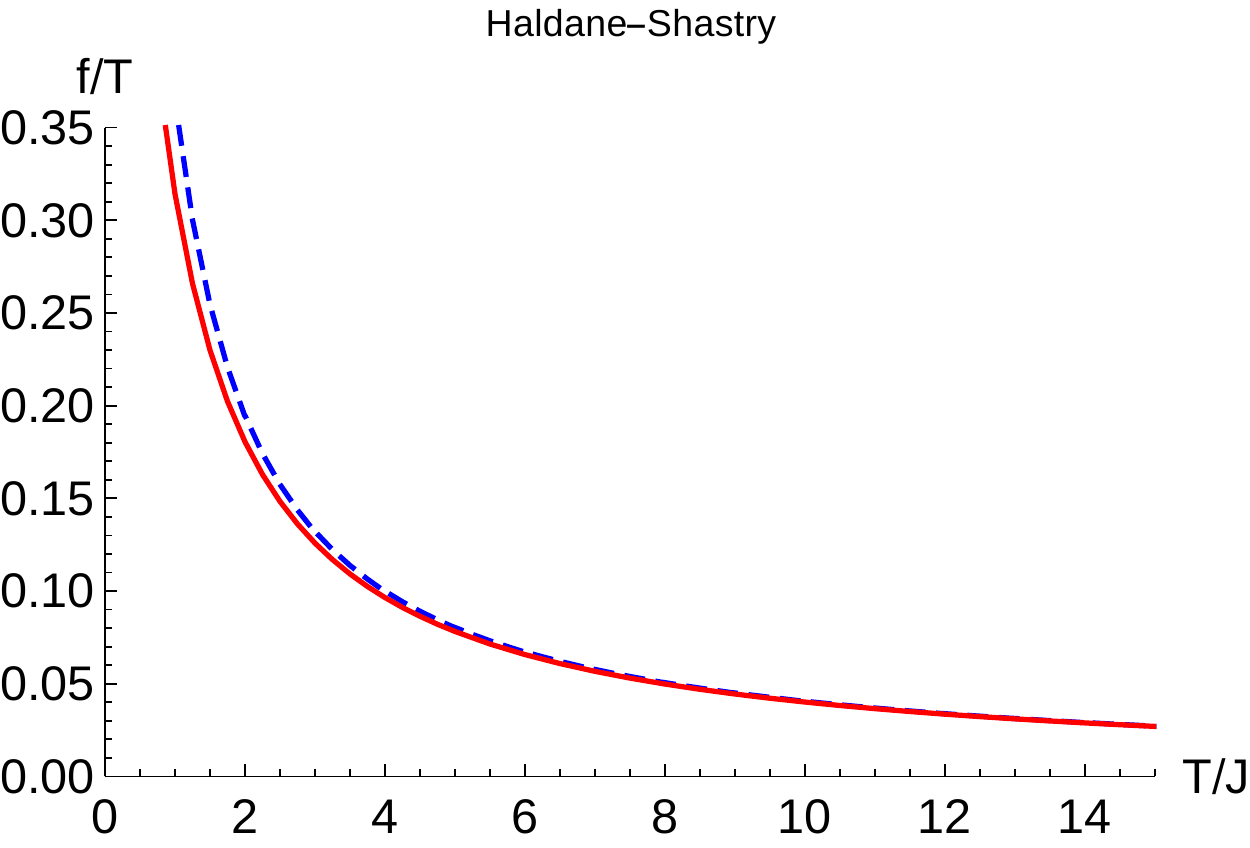}
\caption{}
\end{subfigure}
\begin{subfigure}{8cm}
\includegraphics[width=8cm]{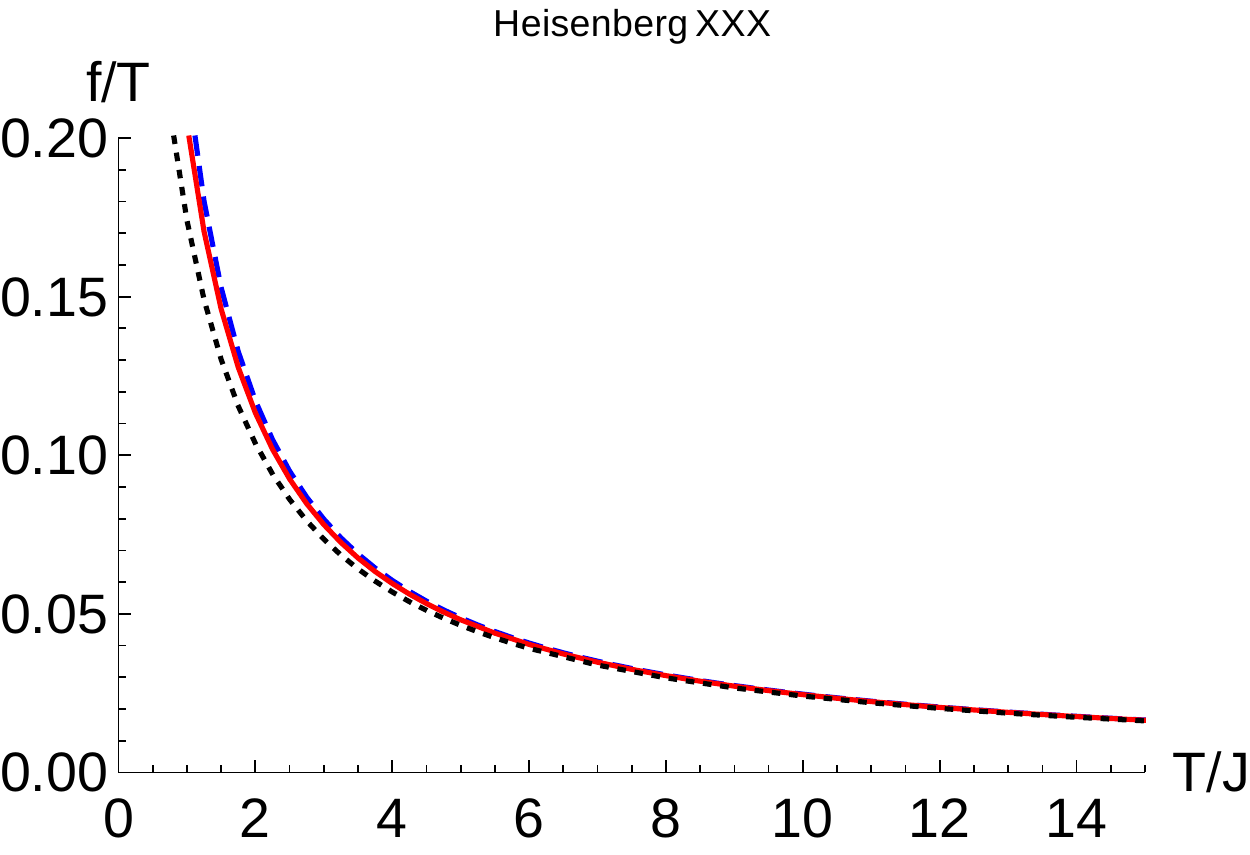}
\caption{}
\end{subfigure}
\\
\begin{subfigure}{\textwidth}
\includegraphics[width=\textwidth]{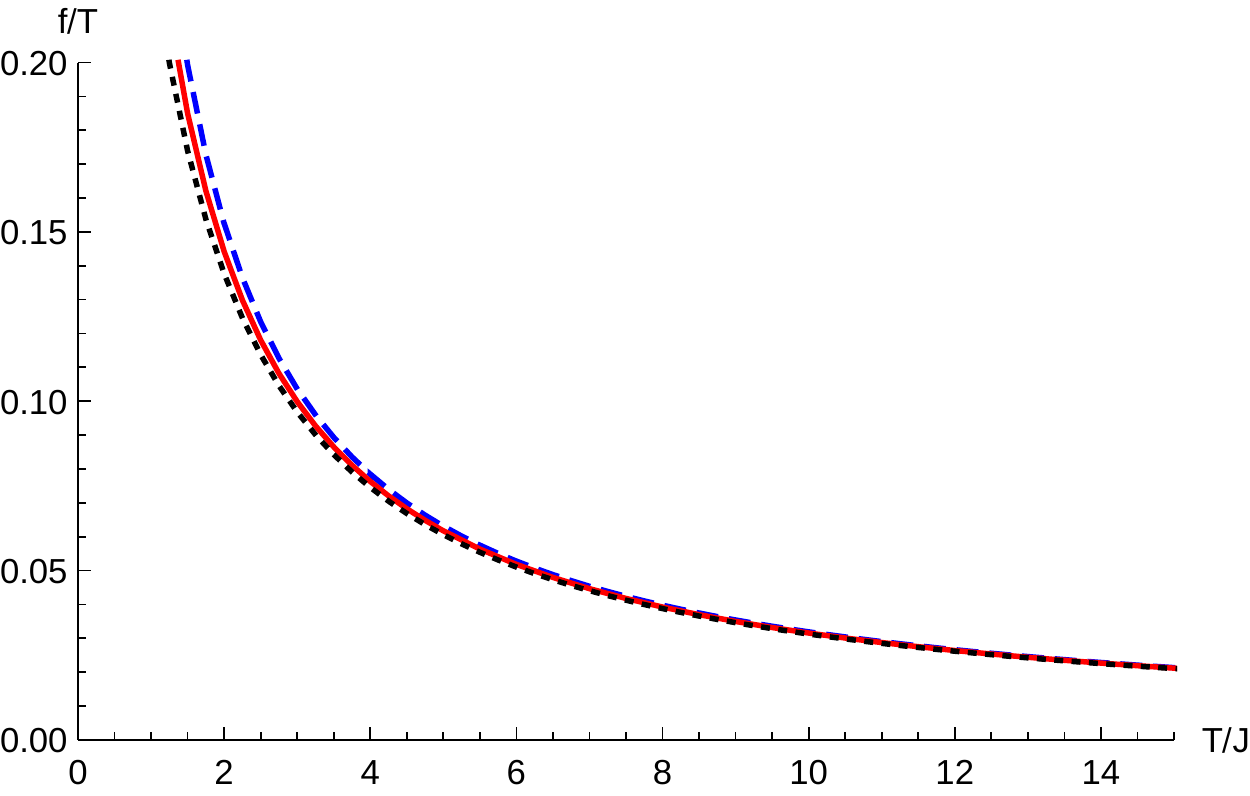}
\caption{}
\end{subfigure}
\caption{(Colour online). Free energies of (a) the Haldane-Shastry spin chain, (b) the Heisenberg \textsc{xxx} spin chain and (c) Inozemtsev's spin chain for $\kappa=1/2$. The blue dashed line is the result from the finite-size hamiltonians using \eqref{matrixformula}, the red solid line is the free energy of the supersymmetric version of the relevant model and the black dotted line is the free energy as calculated from TBA. Note that TBA is absent for the HS spin chain.}
\label{InoSSTr}
\end{figure}

Fig. \ref{InotoXXX} shows the free energy\footnote{All free energies plotted in \ref{InotoXXX} and \ref{InoSSTr} have been offset by $\log 2$, such that they vanish as $T\rightarrow \infty$.} as calculated from the TBA-equations for different $\kappa$, along with the free energy of the Heisenberg \textsc{xxx} spin chain calculated from their TBA-equations and 
(for completeness) the free energy of the Haldane-Shastry spin chain as computed from equation \eqref{matrixformula}. We see that as $\kappa$ increases the free energy of Inozemtsev's spin chain converges to the free energy of the \textsc{xxx} spin chain. Also, for decreasing $\kappa$ the free energy approaches the free energy of the HS spin chain. This shows that our TBA-equations reproduce the thermodynamic behaviour of the two limiting spin chains in the appropriate limits and nicely interpolate between them at finite $\kappa$.
\\[5mm]
In Fig. \ref{InoSSTr} we have plotted free energies of our three models as calculated from \eqref{matrixformula} and the TBA-equations when relevant, accompanied by the free energy of the supersymmetric version of these models. We see that all the free energies agree to very high accuracy for $T\gtrsim 5 J $, whereas deviations occur for smaller $T$. The differences between the different functions scale as $J/T$, as can be confirmed by repeating the analysis for different values of $J$. Moreover, these deviations occur for all our models, including the Heisenberg \textsc{xxx} and Haldane-Shastry spin chains for which previous studies have confirmed the correctness of the underlying equations \cite{finkellopez2,takahashi2005thermodynamics,enciso1}. The deviations are most likely caused by numerical inaccuracies related to exponentiating large numbers (as happens in equation \eqref{matrixformula} and in the TBA-equations \eqref{TBA}) and to restricting the real line to a finite interval. We estimate the observed discrepancies to be within the error of these numerical effects. Therefore, we regard the data in Fig. \ref{InoSSTr} as confirmation that our TBA-equations \eqref{TBA} truly determine the thermodynamic behaviour of Inozemtsev's spin chains. In particular, this validates our usage of the string hypothesis in the derivation of the TBA-equations which is non-trivial in itself. Moreover, the matching with thermodynamic data of the supersymmetric models confirms that the hypothesis brought forward by Finkel and Gonz{\'a}lez-L{\'o}pez in \cite{finkellopez2} that the supersymmetric version coincides with the non-supersymmetric model in the thermodynamic limit. It would be interesting to see whether it is possible to derive the defining equation for the free energy of the supersymmetric model from our TBA-equations \eqref{TBA}, perhaps providing more insight into why this correspondence between certain models and their supersymmetrization exists.  
\\[5mm]
One can further check the claim that Inozemtsev's finite-length model \eqref{inoresc} and infinite-length model \eqref{infIno} coincide in the thermodynamic limit by computing the free energy of the finite-length spin chain given by the hamiltonian \eqref{genham} with potential
\be
\label{infIno2}
V_{\kappa,\infty}^{(L)}(j) = \frac{\sinh^2\kappa}{\sinh^2 \kappa j}
\ee
using \eqref{matrixformula}. The resulting free energy coincides to such high accuracy with the result obtained using the hamiltonian with potential \eqref{inoresc} that they would not be separately discernable in the plots in Fig. \ref{InoSSTr}. This, combined with the fact that the limits to the \textsc{xxx} and HS spin chain behave as expected provide additional evidence that our TBA-equations \eqref{TBA} are correct.
\section{Conclusions}
In this paper we have investigated the thermodynamics of Inozemtsev's elliptic spin chain. Starting from the Bethe ansatz equations of Inozemtsev's infinite-length spin chain, we classified all the solutions to these equations that have real energy and total momentum. We then analyzed whether they parametrize bound-state solutions of Inozemtsev's infinite-length spin chain, for example by comparing the results with one of the limiting cases, the infinite-length Heisenberg \textsc{xxx} spin chain. This reduces the number of solutions immensely, leaving a set of solutions that is structurally very similar to the string solutions of the Bethe ansatz equations of the \textsc{xxx} spin chain. One interesting new feature is the presence of solutions with non-selfconjugate momenta. Carrying out the thermodynamic Bethe ansatz program we have derived a set of coupled integral equations and an associated set of finite-difference equations ($Y$-system) that allows one to compute the free energy of the model at thermal equilibrium. We have solved the integral equations numerically and compared them with the free energy computed directly from the finite-size hamiltonian as well as with the free energy of its limiting models, the Heisenberg \textsc{xxx} and Haldane-Shastry (HS) spin chains. All the results seem to be consistent, corroborating the correctness of our derived integral equations. Moreover, we also compared the free energy of Inozemtsev's spin chain with the free energy of the supersymmetric version of this spin chain obtained by Finkel and Gonz{\'a}lez-L{\'o}pez and concluded that these models coincide in the thermodynamic limit.  
\\
Our findings extends the relationship between Inozemtsev's spin chain and the \textsc{xxx} and HS spin chains to the thermodynamic regime, in line with the finding that this is also true for the supersymmetric version of these models \cite{finkellopez2}. One might wonder whether similar relations exist for other generalizations or deformations of such spin chains. 
Also, further research could be conducted to see whether one can analytically show the equivalence of the normal and supersymmetric Inozemtsev spin chain in the thermodynamic limit as has been done for their limiting models. 
\\
It would be interesting to get a better understanding of the relation between our $Y$-system for Inozemtsev's elliptic spin chain and $Y$-systems for related su$(2)$-invariant models, for example because it might lead to an elliptic extension of the kernel identities we used to derive the $Y$-system. Moreover, it would be interesting to see whether one can simplify the $Y$-system even further in light of the recent advances in simplifying the $Y$-system for $\mathcal{N}=4$ super Yang-Mills theory to what is known as the quantum spectral curve \cite{Gromov2015}. 
\section*{Acknowledgements}
I am indebted to G. Arutyunov for bringing Inozemtsev's spin chains to my attention. I would also like to thank him and W. Galleas, J. Lamers, J. van de Leur and S. van Tongeren for useful discussions and G. Arutyunov, W. Galleas and J. Lamers for comments on the manuscript. This work is supported by the German Science Foundation (DFG) under the Collaborative Research Center (SFB) 676 ''Particles, Strings and the Early Universe'' and the Research Training Group (RTG) 1670 ''Mathematics inspired by String Theory and Quantum Field Theory''. 
\newpage
\begin{appendices}
\addtocontents{toc}{\setcounter{tocdepth}{-1}}
\section{Properties of Weierstra{\ss} Elliptic Functions}
\label{sec:elliptic}
The Weierstra{\ss} elliptic functions are defined using a lattice $\mathbb{L}$ that defines the periodicity of these functions (see for example \cite{Whittaker,dlmf}):
\be
\mathbb{L} := \{ z\in \C | z= n\omega_1 +m\omega_2, n,m\in \Z\},
\ee
where the $\omega_i$ are the periods of the lattice and obey Im$(\omega_1/\omega_2)<0$. The definitions of the Weierstra{\ss} elliptic functions can now be written as
\bea
\wp(z) &=& \frac{1}{z^2} + \sum_{\substack{\omega \in \mathbb{L} \\ \omega\neq 0}} \left(\frac{1}{(z-\omega)^2} -\frac{1}{\omega^2}\right) \nonumber \\
\zeta(z) &=&  \frac{1}{z} + \sum_{\substack{\omega \in \mathbb{L} \\ \omega\neq 0}} \left(\frac{1}{z-\omega} +\frac{1}{\omega}+\frac{z}{\omega^2}\right),
\eea
where all these series converge absolutely and uniformly for $z \in A \subset \C$ for all compact $A$ satisfying $A\cap \mathbb{L} = \emptyset$. Moreover, $\wp$ is even and meromorphic with double poles with residue $0$. $\zeta$ is odd and meromorphic with simple poles with residue $1$. Note that formally $\zeta$ is not doubly periodic and hence not elliptic.
\\[5mm]
These functions furthermore satisfy
\bea
\wp(z)&=& -\zeta'(z)
\eea
for all $z \not \in \mathbb{L}$.
\section{Behaviour of $\phi$}
\label{sec:B}
\begin{wrapfigure}{r}{0.4\textwidth}
  \vspace{-20pt}
\begin{center}
\begin{tikzpicture}[>=latex,scale=0.6]
\path [fill=MyColorOneLightest] (0,-3) rectangle (6.28,3);
\fill [MyColorOneLightest] (0,0) circle (0.2);
\fill [white] (6.28,0) circle (0.2);
\draw[thick] (-1,0) -- (7,0);
\draw[thick, dashed] (0,3) -- (6.28,3);
\draw[thick, dashed] (0,-3) -- (6.28,-3);
\draw[thick] (3.14,-5) -- (3.14,5);
\draw[thick, dashed] (-0.02,3) -- (-0.02,0.2);
\draw[thick, dashed] (-0.02,-3) -- (-0.02,-0.2);
\draw[thick, dashed](-0.02, 0.2) arc (90:270:0.2);

\draw[thick, dashed] (6.26,3) -- (6.26,0.2);
\draw[thick, dashed] (6.26,-3) -- (6.26,-0.2);
\draw[thick, dashed](6.26, 0.2) arc (90:270:0.2);

\node[above] at (2.5,-3.8) {$-\kappa i$};
\node[above] at (2.7,3) {$\kappa i$};
\node[above] at (6.6,0) {$\pi$};
\node[above] at (-0.6,0) {$-\pi$};
\node[above] at (2.8,0) {$0$};
\fill [black] (3.14,0) circle (2pt);
\fill [black] (0,0) circle (2pt);
\fill [black] (6.28,0) circle (2pt);
\fill [black] (3.14,3) circle (2pt);
\fill [black] (3.14,-3) circle (2pt);
\node[below] at (7.3,0) {Re$(p)$};
\node[left] at (3,5) {Im$(p)$};

\end{tikzpicture}
\end{center}
\caption{The contour around which we integrate to find the number of zeroes in $D_{{\scriptscriptstyle \leq}\kappa}$ for $\phi$.}
  \vspace{0pt}
\label{contour}
\end{wrapfigure}
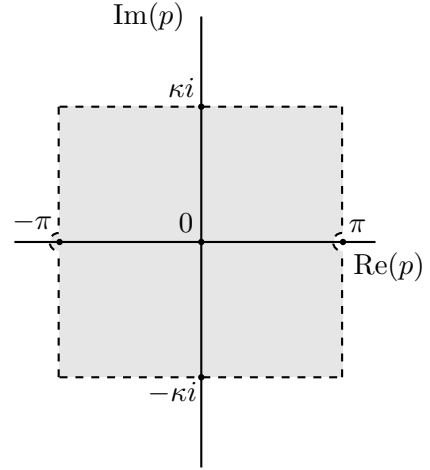

We investigate the behaviour of $\phi$ on the region $D_{{\scriptscriptstyle \leq}\kappa}$ \eqref{Dsmallkappa}. Consider the contour $C$ depicted in Figure \ref{contour}, which travels around $D_{{\scriptscriptstyle \leq}\kappa}$ counterclockwise on the boundary. In its interior, there is one pole, at $z=0$. Note that due to the periodicity of $\phi$ in the real direction, the small deviation around the the points $\pm \pi$ does not affect the analysis.
\\[5mm]
We can also find the imaginary part of $\phi$ on the top and bottom edge of this contour by a simple observation: let $x \in \R$, then $\overline{\phi(x-i\kappa)}=\phi(x+i\kappa) = \phi(x-i\kappa)-i$ by quasi-periodicity and we have\footnote{ $\overline{\phi(z)} = \phi\left(\overline{z}\right)$ follows from the oddity of $\zeta$ in the definition of $\phi$.}
\bd
\overline{\phi(x-i\kappa)}-\phi(x-i\kappa) = i
\ed
which implies that Im$\left(\phi(x-i\kappa)\right) = -i/2$ and Im$\left(\phi(x+i\kappa)\right) = i/2$. Thus on the top and bottom 
edge of this contour, the imaginary part of $\phi$ is constant. Let $\theta \in \mathbb{C}$ be arbitrary, but such that $\phi(z) = \theta$ has no solutions when $\phi$ is restricted to the contour. Then the function $\tilde{\phi}(z) = \phi(z)-\theta$ has no zeroes or poles on the contour and we can use the argument principle to state that
\be
\oint_C \frac{\tilde{\phi}'(z)}{\tilde{\phi}(z)} dz = 2\pi i \left( N -P \right),
\ee
where $N$ is the number of zeroes and $P$ the number of poles of $\tilde{\phi}$ in the interior of the contour, which is the fundamental region of $\phi$. In this case, we have $P=1$. We can calculate the integral on the left-hand side: the contributions from the vertical parts of the contour cancel each other due to the periodicity of $\tilde{\phi}$. For the contributions of the top part, we see the following: 
\be
\int_{-\pi}^{\pi} \frac{\tilde{\phi}'(x+\kappa i)}{\tilde{\phi}(x+ \kappa i)} dx = \int_{-\pi}^{\pi} \frac{d}{dx}\log\left(\tilde{\phi}(x+ \kappa i)\right) dx = \log\left(\tilde{\phi}(\kappa i)\right)-\log\left(\tilde{\phi}(2\pi +\kappa i)\right) = 0,
\ee
because $\tilde{\phi}$ is $2\pi$-periodic in the real direction. Note that we could evaluate the integral using the logarithm, because we know that the imaginary part of  $\tilde{\phi}$ is constant along the path, allowing us to find a holomorphic branch for the logarithm on a neighbourhood of the top part of the contour. In a similar fashion, one can show that the contribution from the bottom part vanishes, thus we end up with 
\bd
\oint_C \frac{\tilde{\phi}'(z)}{\tilde{\phi}(z)} dz = 0,
\ed
implying that for all the $\theta$ we considered, $\tilde{\phi}$ has exactly one zero in the fundamental region, thus $\phi(z) = \theta$ has exactly one solution in this region. 
\\[5mm]
On the boundary of $D_{{\scriptscriptstyle \leq}\kappa}$, the following holds. The restriction 
$x\mapsto \phi(- \pi + ix)$ (with $x\in [-\kappa,\kappa]$) has negative derivative everywhere. Moreover, since $\phi(- \pi \pm \kappa i) = \mp i/2$, we can conclude that this restriction maps bijectively onto $[-i/2,i/2]$. This shows that $\phi: [-\pi,\pi[\,\,\oplus\,\,]{-}i \kappa,i \kappa[\,\rightarrow A\subset\C$ maps bijectively onto its image $A$. On the top part of the contour we can write $x\mapsto \phi(x+i\kappa)$ for the restriction. A plot of this function is shown in Figure \ref{phibound}, which shows that this restriction is not bijective onto its image. In fact, all image values are attained exactly twice. We call the graphs maximum $\phic$ and by symmetry, its minimum is $-\phic$. The value of $p$ for which Re($\phi(p+i\kappa)) = \phic$ we call $\pc$. By symmetry, the minimum is attained at $-\pc$. The behaviour of the real part of $\phi$ along the bottom boundary is exactly the same.
\\[5mm]
We can now conclude that $\phi$ is surjective onto $\C$ and almost injective: the only values it attains twice are those of the form $\theta \pm i/2$, where $|\theta| \leq \phic$. 
\section{Convergence of coinciding solutions}
\label{sec:C}
\subsection{General analysis}
The addition of extra signs $s_{j,{i_j}}$ to a basic string solution such as in Fig. \ref{tree1} seems, at least at first glance, to work fine together with the reasoning we employed before: for extra $s_{j,{i_j}}$ associated to $\theta_j$ anywhere on the string except for the endpoints, there are $s_n$ in the set such that there is a term on the right-hand side of the Bethe ansatz equations (BE) which have the right convergence properties. Conform the main text we will call these signs \emph{helping}. However, usually little attention is given to the fact that precisely due to the simple structure of the strings in $\theta$-space there are almost always terms in the BE which have the opposite convergence behaviour. We call the associated signs \emph{counteracting signs}. To address this issue more precisely, we must take a closer look at what happens when taking the limit $L\rightarrow \infty$. We therefore associate to each $p_j$ in a coinciding solution a $\delta_j>0$ that indicates how fast the solution converges in the limit $L\rightarrow \infty$ in the following sense: we associate to each $p_j$ a sequence $\left(p_j^{(L)}\right)$ indexed by $L$ with limit $p_j$, which gives rise to a sequence in $\theta$-space $\left(\phi\left(p_j^{(L)}\right)\right)$ with limit $\theta_j = \phi(p_j)$. Since the left-hand side of the Bethe equations converges to $0$ (or diverges to infinity) exponentially, the right-hand side should do the same, implying that the image point sequences should converge exponentially. We define $\delta_j$ such that for large $L$
\bd
\left| \phi\left(p_j^{(L)}\right) - \phi(p_j)\right|=\mathcal{O}\left(e^{-\delta_j L}\right).
\ed
Let us now consider the convergence properties of the BE of a momentum $p_{j,{i_j}}$ associated to the sign $s_{j,{i_j}}=+$ and such that Im$(p_{j,{i_j}}) = \theta_I+j-1$ sitting on level $j$. In the limit $L\rightarrow \infty$, the BE associated to a sign $s_{j,{i_j}}=+$ is satisfied if the right-hand side goes to $0$, which is achieved by the existence of signs on level $j+1$, cf. the discussion in Section \ref{sec:treestructures}. On the other hand, the terms on the right-hand side of the Bethe equation associated to the signs on level $j-1$ go to infinity, they are counteracting. It seems that the right-hand side of the Bethe equation has the right limit only if the terms associated to helping signs converge faster than those associated to the counteracting signs. However, for minus signs the situation is exactly opposite: the signs on level $j-1$ are helping, those on level $j+1$ are countaracting. These two observations seem to contradict each other, but this is not completely true, as the following analysis will show.
\\[5mm]
Define the convergence rates of $p_{j,\alpha_j}$ (with positive imaginary part) as $\delta_{j,\alpha_j}$ and of $p_{j,\beta_j}$ (with negative imaginary part) as $\delta_{j,\beta_j}$. Consider the $n_j$th plus sign on level $j$ in a tree solution. The Bethe equation of the momentum associated to this plus sign reads
\be
\label{beq2}
e^{ip_{j,n_j}L} = \prod_{\substack{k=1 \\ k\neq j}}^M \left( \frac{\theta_{j,n_j} - \theta_k+i}{\theta_{j,n_j} - \theta_k -i}\right)^{(P_k+M_k)},
\ee
where we have $\theta_{j,n_j} =\phi(p_{j,n_j})$ and $\theta_k$ belongs to level $k$. Note that the terms belonging to other momenta on level $j$ are all $1$ and are not written explicitly and that the $\theta_{j,n_j}$ do not actually depend on $n_j$. As $L\rightarrow \infty$, the left-hand side converges to $0$. Most of the terms on the right-hand side converge to finite values and are irrelevant for the analysis. The interesting terms are those belonging to level $j \pm 1$. They form the product
\be
\label{beq3} 
\underbrace{\frac{\theta_{j,n_j} - \theta_{j+1}+i}{\theta_{j,n_j} - \theta_{j+1} -i}\cdots\frac{\theta_{j,n_j} - \theta_{j+1}+i}{\theta_{j,n_j} - \theta_{j+1} -i}}_{P_{j+1}+M_{j+1}}
\underbrace{\frac{\theta_{j,n_j} - \theta_{j-1}+i}{\theta_{j,n_j} - \theta_{j-1} -i}\cdots\frac{\theta_{j,n_j} - \theta_{j-1}+i}{\theta_{j,n_j}- \theta_{j-1} -i}}_{P_{j-1}+M_{j-1}}.
\ee
However, to each momentum we have associated a convergence rate and we can let all the fractions in this product converge to their limiting value with different rates. In the infinite-$L$ limit, the term belonging to $p_{j+1,\gamma_{j+1}}$ (with $\gamma = \alpha,\beta$) on the level $j+1$ behaves as
\be
\left| \frac{\theta_{j,n_j} - \theta_{j+1} +i}{\theta_{j,n_j} - \theta_{j+1} -i} \right| \approx 
\mathcal{O}\left(\exp\left[-\min\left( \delta_{j,n_j},\delta_{j+1,\gamma_{j+1}}         \right)L\right]\right),
\ee
while the term belonging to $p_{j-1,\gamma_{j-1}}$ behaves as 
\be
\left|\frac{\theta_{j,n_j} - \theta_{j-1} +i}{\theta_{j,n_j} - \theta_{j-1} -i} \right| \approx 
\mathcal{O}\left(\exp\left[{\min\left( \delta_{j,n_j},\delta_{j-1,\gamma_{j-1}}         \right)}L\right]\right).
\ee
From now on, we write $(x,y):= \min(x,y)$. In total, the product of terms belonging to level $j+1$ converges as 
\bd
\mathcal{O}\left( \exp\left[ -  \sum_{\alpha_{j+1} =1}^{P_{j+1}} \left( \delta_{j,n_j},\delta_{j+1,\alpha_{j+1}}\right) -  \sum_{\beta_{j+1} =1}^{M_{j+1}} \left( \delta_{j,n_j},\delta_{j+1,\beta_{j+1}}\right) \right] \right)
\ed
and combining this with the similar result for the level $j-1$ we see that the right-hand side of the Bethe equation \eqref{beq2} behaves as 
\bea
\mathcal{O}\left( \exp\left[ -  \sum_{\alpha_{j+1} =1}^{P_{j+1}} \left( \delta_{j,n_j},\delta_{j+1,\alpha_{j+1}}\right) -  \sum_{\beta_{j+1} =1}^{M_{j+1}} \left( \delta_{j,n_j},\delta_{j+1,\beta_{j+1}}\right) \right.\right. \nonumber \\
\left.\left.  + \sum_{\alpha_{j-1} =1}^{P_{j-1}} \left( \delta_{j,n_j},\delta_{j-1,\alpha_{j-1}}\right) + \sum_{\beta_{j-1} =1}^{M_{j-1}} \left( \delta_{j,n_j},\delta_{j-1,\beta_{j-1}}\right)
 \right] \right)
\eea
and therefore goes to zero only when the convergence rates obey
\bea
\label{rest1}
-  \sum_{\alpha_{j+1} =1}^{P_{j+1}} \left( \delta_{j,n_j},\delta_{j+1,\alpha_{j+1}}\right) -  \sum_{\beta_{j+1}=1}^{M_{j+1}} \left( \delta_{j,n_j},\delta_{j+1,\beta_{j+1}}\right)  \nonumber \\ + \sum_{\alpha_{j-1} =1}^{P_{j-1}} \left( \delta_{j,n_j},\delta_{j-1,\alpha_{j-1}}\right) + \sum_{\beta_{j-1} =1}^{M_{j-1}} \left( \delta_{j,n_j},\delta_{j-1,\beta_{j-1}}\right) <0.
\eea
In a similar fashion, one can derive that the Bethe equation corresponding to a momentum $p_{j,n_j}$ with negative imaginary part is satisfied only when 
\bea
\label{rest2}
-  \sum_{\alpha_{j+1} =1}^{P_{j+1}} \left( \delta_{j,n_j},\delta_{j+1,\alpha_{j+1}}\right) -  \sum_{\beta_{j+1} =1}^{M_{j+1}} \left( \delta_{j,n_j},\delta_{j+1,\beta_{j+1}}\right)  \nonumber \\  + \sum_{\alpha_{j-1} =1}^{P_{j-1}} \left( \delta_{j,n_j},\delta_{j-1,\alpha_{j-1}}\right) + \sum_{\beta_{j-1} =1}^{M_{j-1}} \left( \delta_{j,n_j},\delta_{j-1,\beta_{j-1}}\right) >0.
\eea
For a valid solution of the Bethe equations, equation \eqref{rest1} must be satisfied for all plus signs, while equation \eqref{rest2} must be satisfied for all minus signs. Note that these restrictions arise simply because there is more than one term that exhibits vanishing or divergent behaviour and we should include more information to find the behaviour of the product. This problem already exists in many of the previously known cases (such as the Hubbard model or the Heisenberg \textsc{xxx} model), but as far as we know, this has never been addressed. Fortunately, however, the restrictions \eqref{rest1},\eqref{rest2} simplify drastically for the usual simple string solutions occurring in the aforementioned cases and can easily be solved. The system of restrictions for a string solution without a real momentum involved read
\bea
(\delta_2^{(+)},\delta_1^{(+)}) - (\delta_2^{(+)},\delta_3^{(+)}) <0 \nonumber \\
\vdots \nonumber \\
(\delta_{m_p}^{(+)},\delta_{m_p-1}^{(+)}) - (\delta_{m_p}^{(+)},\delta_{m_p+1}^{(-)}) <0 \nonumber \\
(\delta_{m_p+1}^{(-)},\delta_{m_p}^{(-)}) - (\delta_{m_p+1}^{(-)},\delta_{m_p+2}^{(-)}) >0 \nonumber \\
\vdots \nonumber \\
(\delta_{M-1}^{(-)},\delta_{M-2}^{(-)}) - (\delta_{M-1}^{(-)},\delta_{M}^{(-)}) >0,
\eea
where the superscripts indicate the sign of the imaginary part of the associated momenta. It is solved by the ordering
\be
\delta_1 <\delta_2 < \cdots <\delta_{m_p} = \delta_{m_p+1} > \delta_{m_p+2} > \cdots >\delta_M,
\ee
together with $\delta_j:=\delta_j^{+}=\delta_j^{-}$. However, determining whether the system of equations consisting of \eqref{rest1} and \eqref{rest2} for a general tree solution can be solved is a much more complicated question. In the next section, we treat some cases and include an example from which it follows that not every sign configuration has a consistent set of convergence rates. 
\subsection{Examples}
A coinciding solution consists of at least $2$ levels. The $2$-level case (illustrated in subfigure (a) in Figure \ref{2level}) can also be solved in general, because the inequalities are trivially satisfied.
\begin{figure}[h]
\begin{center}
\begin{subfigure}{0.3\textwidth}
\begin{tikzpicture}
\node at (1.5,0) {$-$};
\node at (2.5,0) {$-$};
\node at (3.2,0) {$\cdots$};
\node at (3.8,0) {$\cdots$};
\node at (4.5,0) {$-$};

\node at (1.5,-1) {$+$};
\node at (2.5,-1) {$+$};
\node at (3.2,-1) {$\cdots$};
\node at (3.8,-1) {$\cdots$};
\node at (4.5,-1) {$+$};

\node at (0,0) {$M_2>0$};
\node at (0,-1) {$P_{1}>0$};
\end{tikzpicture}
\caption{}
\end{subfigure} 
\qquad
\begin{subfigure}{0.3\textwidth}
\centering
\begin{tikzpicture}
\node at (1.5,1) {$-$};

\node at (1,0) {$+$};
\node at (2,0) {$-$};
\node at (1.5,-1) {$+$};

\end{tikzpicture}
\caption{}
\end{subfigure}
\end{center}
\caption{(a) A $2$-level coinciding solution. (b) A $3$-level coinciding solution that does not admit a consistent set of convergence rates.}
\label{2level}
\end{figure}
However, already the $3$-level case harbours an example of a configuration that cannot have a consistent set of convergence rates. Consider the example in subfigure (b) in Figure \ref{2level}. The relevant set of equations is 
\bea
\label{ex1}
(\delta_2^{(+)},\delta_1) - (\delta_2^{(+)},\delta_3) <0 \nonumber \\
(\delta_2^{(-)},\delta_1) - (\delta_2^{(-)},\delta_3)>0,
\eea
where we omit the superscript $(\pm)$ when it is not necessary. We first try to deduce which of the $\delta$'s should be the smallest one of these four. From the upper equation, we conclude that neither $\delta_2^{(+)}$ nor $\delta_3$ can be the smallest, while the lower equation tells us that neither $\delta_2^{(-)}$ nor $\delta_1$ can be the smallest. Therefore, none of the $4$ rates can be the smallest, thus no solution can exist. Note that this example can be extended: if we include $P_2>0$ pluses and $M_2>0$ minuses on level $2$, the resulting set of restrictions has the system \eqref{ex1} as a subsystem and cannot be solved. In particular, this shows that example (b) we treated in Section \ref{sec:treestructures} is not a valid solution to the BE after all, although we could find momenta to match the configuration. Moreover, any sign configuration that contains this $3$-level structure cannot be solved. However, all other $3$-level configurations do admit a consistent solution as a careful analysis of the cases shows.
\\[5mm]
We have not been able to find a general algorithm to solve these complex coupled sets of inequalities or prove the existence (or absence) of a solution. The only configurations we found that lead to inconsistent inequalities are of the type described in the previous paragraph. In any case, the structure of the solutions is complicated, but in the present analysis we do not need it. 
\addtocontents{toc}{\setcounter{tocdepth}{2}}
\end{appendices} 
\newpage
\bibliography{bibliography3}
\addcontentsline{toc}{section}{References}
\end{document}